\documentclass{aa}
\usepackage{graphicx}
\newcommand{\msun}{M$_{\odot}$}
\begin{document}

%

   \title{Dynamics of blue compact galaxies, \\
 as revealed by their H$\alpha$ velocity fields \thanks{Based on observations collected 
at the European Southern Observatory, La Silla, Chile; 
}
}

\subtitle{II. Mass models and the starburst triggering mechanism}

\author{G\"oran \"Ostlin \inst{1}
\and Philippe Amram \inst{2} 
\and Nils Bergvall \inst{3}  
\and Josefa Masegosa \inst{4} 
\and Jaques Boulesteix \inst{2}
\and Isabel M\'arquez \inst{4}
}
 
\offprints{G\"oran \"Ostlin, ostlin@astro.su.se}

\institute{ Stockholm Observatory, SE-133~36 Saltsj\"obaden, Sweden
\and
Laboratoire d'Astrophysique de Marseille, Observatoire de Marseille, 2 Place le Verrier, F-13248 Marseille Cedex 04, France
\and
Astronomiska observatoriet, Box 515, S-75120 Uppsala, Sweden
\and
Instituto de Astrof\'{\i}sica de Andaluc\'{\i}a (CSIC), Apdo. 3004, E-18080 Granada, Spain
}

\date{Received 15 January 2001, Accepted 17 May 2001 }

\authorrunning{\"Ostlin et al.}
\titlerunning{Dynamics of blue compact galaxies II.}

   \abstract{ 
The H$\alpha$ velocity fields of a sample of six
luminous blue compact galaxies (BCGs) and two companions have been
obtained by observations with a scanning Fabry-Perot interferometer.
The Fabry-Perot images, velocity fields and rotations curves have been
presented in a previous paper (Paper I). In general, the velocity
fields are irregular and often contain secondary dynamical components,
but display overall rotation. The two companions have more regular
velocity fields and rotation curves.
In this article we analyse the velocity fields and dynamics together
with the morphology of the studied BCGs, and present detailed mass
models. In addition, we model the stellar mass content by means of
multicolour surface photometry and spectral evolutionary synthesis analysis. By
comparison of the masses of stars and those derived from the rotation
curve, we show that about half of the galaxies cannot be supported by
rotation alone. The morphology and dynamics of the BCGs suggest that the
starburst activity in these galaxies are most likely triggered by
mergers involving gas-rich dwarf galaxies and/or massive gas clouds.
}

   \maketitle

\section{Introduction}

The blue compact galaxies (BCGs) are characterised by their blue
colour, strong nebular emission lines, compact appearance and low
chemical abundances (e.g. Searle and Sargent 1972, Lequeux et
al. 1979, Kunth and Sargent 1983; Masegosa et al. 1994, Izotov and 
Thuan 1999). The low
chemical abundances and high star formation rates (SFRs) opened up the 
possibility that these were  genuinely young galaxies presently forming 
their first generation of stars (Searle and Sargent 1972). Since 
these BCGs were found at low red-shift, this would mean that galaxies 
could still be forming in the local universe. If that would be the 
case, it would have considerable implications on theories on galaxy and
structure formation, and on the nature of dark matter.

Since this was proposed, many BCGs have been found to contain a
population of old stars. Evidence comes mainly from the studies of
colours and morphologies of the faint halos seen around most BCGs (Loose
and Thuan 1985; Kunth et al. 1988; Papaderos et al. 1996; Telles and 
Terlevich 1997; Doublier et al. 1999;
Bergvall and \"Ostlin 2000).  Regular red halos are taken as evidence
that there is an underlying old population present. Despite a few
remaining young galaxy candidates it is evident that most BCGs are in
fact old (see Kunth and \"Ostlin 2000 for a review). What is less
evident, however, is the reason why many BCGs seem to be forming stars 
at such high rates. 

For many BCGs, the gas consumption time scales are significantly shorter 
than a Hubble time  (e.g. Fanelli et al. 1988), which indicates that the 
current high SFRs have to be transient.
The general picture has been  that these galaxies undergo a few or several
short bursts of star formation followed by longer more quiescent
periods (Searle et al. 1973, Gerola et al. 1980). The explanation for
this could be gas dynamical; the gravitational field has to compete
with galactic winds produced by mass loss and supernovae (SNe),
which may expel the gas and prevent further star formation. Depending
on the mass of the galaxy and the pressure in the surrounding
intergalactic medium, the gas may cool and later accrete back on the
galaxy producing a new burst (Babul and Rees 1992). It is believed
though that the number of bursts has to be few, from gas
consumption arguments and in order not to exceed the observed
metallicities.  However, the latter argument is not so strong in view
of the possibility that BCGs lose enriched gas in supernovae winds
created shortly after the onset of the starburst. On the other hand,
as winds travel through the galaxy, the newly synthesised metals will
mix with the surrounding ISM. Thus an outflow of gas is not
necessarily more metal rich than the ISM in the galaxy (Ferrara and
Tolstoy 2000).
 
Other explanations for the starburst phenomenon involve tidally triggered 
gas infall through interactions with companion galaxies (Lacey and Silk 1991).
In this case one could imagine a disturbed gas-rich dwarf  galaxy, or even 
an unevolved protogalactic cloud, as the progenitor of a BCG.  
However, some investigations show that most BCGs are fairly
isolated: Campos-Aguilar et al. (\cite{campos:1993})  investigated the 
environments of H{\sc ii}-galaxies from the spectrophotometric catalogue 
by Terlevich et al. (\cite{terlevich:etal}) by comparing their location 
in space with those of galaxies in the CfA catalogue. They found that most 
H{\sc ii}-galaxies were isolated, and those that were not still had an 
average distance to its nearest neighbour of several hundreds of kpc. 
Similar results were obtained by Telles and Terlevich (\cite{telles:1995}).
Salzer (\cite{salzer}) found that emission line galaxies were less clustered
then normal giant galaxies and that they tended to avoid 
regions with high galactic density (of massive galaxies), similar to
the results obtained for UV excess galaxies by Iovino et al. (1986),
and recently by Telles \& Maddox (2000).  

On the other hand, Taylor et al. (\cite{taylor:1995},\cite{taylor:1996a}, 
erratum\cite{taylor:1996b}; Taylor 1997) made H{\sc i} surveys for companion 
galaxies around  H{\sc ii}-galaxies and LSBGs.  They 
found that nearly 60\% of the H{\sc ii}-galaxies  in their sample had H{\sc i} 
companions.  The important difference with respect to the studies above is that 
this survey was a direct search sensitive also to faint gas-rich dwarf galaxies, 
while the others used
catalogues with bad completeness for dwarfs. Thus, while BCGs  tend 
to avoid giant luminous galaxies, they appear to have dim low-mass 
neighbours in many cases.

In any case, it is likely that the BCGs hold some key information about 
the early evolution and formation of dwarf and intermediate sized galaxies. 
If it can be understood how BCGs evolve on a longer time scale and under 
what circumstances an 'object' can be transformed into a BCG, much will 
have been learned about the evolution of galaxies. 
Some have  proposed a relationship with faint blue galaxies (FBGs) at 
z$\sim$0.5 (Cowie et al. 1991; Babul and Rees 1992), others with low surface
brightness galaxies (LSBGs), which may be the precursors and/or the successors 
of BCGs (Bergvall et al. 1999a, 1999b; Telles and Terlevich 1997). Further, an 
understanding of BCGs may help in interpreting observations of star forming 
objects at high red-shifts. The studies of intermediate--high red-shift compact 
galaxies in the  Hubble deep field (Guzm\'an et al. 1996, 1997; Phillips et
al. 1997) reveal that these include galaxies with properties very similar 
to those of local luminous BCGs.

Fully understanding the evolution of BCGs requires detailed knowledge 
about their star formation histories (SFHs),  dynamics, and possible 
merger history. Most BCGs are far too distant to be resolved into stars
even with the Hubble Space Telescope (HST). Instead, the SFHs can be examined 
with the help of spectral evolutionary synthesis models (SEMs) and 
optical surface photometry (\"Ostlin and Bergvall 1994,
\"Ostlin et al.  1996, Bergvall and \"Ostlin 2000).  This can put
constraints both on star formation processes, including the initial
mass function (IMF) of stars and the number of old stars present.  
The tricky part is to constrain the mass of the old, cool stellar component,
since the light in the optical is totally dominated by hot/massive
stars.  The old population is easier approached in the halos, outside
the starburst region, and by adding observations in the near infrared
(NIR) (Bergvall and \"Ostlin 2000).   A lot could also be gained with 
knowledge of the dynamics of BCGs, from which one can then put interesting 
constraints on the amount of old stars and dark matter present, and
provide important information on the triggering mechanism behind starbursts 
in BCGs.

Previous studies of the dynamics of BCGs include optical and radio
investigations.  The latter target the 21cm emission from neutral
hydrogen and is very useful for studying the large scale dynamics,
while the inferior spatial resolution (as compared to the optical
regime) make them of limited use for studying the central dynamics and
the kinematics of the star forming regions. Optical studies have
probed either the full two dimensional velocity field, e.g. by
utilising Fabry-Perot interferometry (Thuan et al. 1987, Marlowe et
al. 1995, Petrosian et al. 1997, \"Ostlin et al. 1999) or limited
parts from slit spectroscopy (e.g. Gil de Paz et al. 1999). The general
picture is that of large scale rotation (\"Ostlin et al. 1999) and
smaller scale distortions e.g. in the form of expanding bubbles
(Marlowe et al. 1995, Martin 1998, Kunth et al. 1998, \"Ostlin et
al. 1999, Gil de Paz et al. 1999). Radio observations provide further
evidence for large scale rotation in BCGs (Viallefond et al. 1987;
Bergvall and J\"ors\"ater 1988; Meurer et al. 1996, 1998; van Zee et
al 1998).

The notion ``starburst'' is frequently   used in the literature to describe 
galaxies, or regions of galaxies, with varying degrees of  star formation  
activity. In this paper, we will adopt a more strict definition by requiring 
that a starburst involves a global SFR which is unsustainably high. The SFR
may be unsustainable because the gas consumption time scale, or the time scale to 
build up the observed stellar mass, is significantly shorter than the Hubble 
time, i.e. the time averaged SFR is an order of magnitude lower than the present. 
Many galaxies 
have SFRs fluctuating with time, and/or have impressive bright H{\sc ii} regions, 
but this does not necessarily imply that the SFR is unsustainable over a 
Hubble time. Indeed, many galaxies classified as BCGs in the literature are 
{\em not} starburst according to our definition.

\subsection{Sample selection and Fabry-Perot observations}

In this paper we will  discuss the results from a study of the 
dynamics of BCGs. In 1995, we obtained  Fabry-Perot interferometric 
observations, targeting the H$\alpha$ emission line, for a sample of 
six luminous BCGs ($M_B \in [-17,-20]$) and two companion galaxies. 
Four of the BCGs were taken from the catalogue of compact galaxies by 
Bergvall and Olofsson (1986). One galaxy (\object{ESO 185-13}) was 
taken from an extension of this catalogue (Bergvall et al. unpublished). 
The last BCG (\object{Tololo 0341-407}) was taken from the spectrophotometric
catalogue of H{\sc ii} galaxies by Terlevich et al. (1991). The galaxies 
were selected to be actively star forming, as judged by the equivalent 
width and luminosity of H$\beta$ or H$\alpha$. The four galaxies from 
the Bergvall and Olofsson (1986) catalogue are the ones with the highest 
the equivalent widths in H$\beta$ from that list. The two other BCGs 
were in addition chosen because they had the right coordinates  with respect 
to the allocated observing time. The reason for choosing intrinsically 
luminous BCGs was to get a more homogeneous sample, in view of the 
fairly small number of galaxies observed. 
Two of the BCGs selected from the Bergvall and Olofsson (1986) catalogue
have confirmed physical star forming companions, and these were also 
included in the target list.

The observations and reductions are thoroughly described 
in Paper I (\"Ostlin et al. 1999, hereafter Paper I), where we also 
present the velocity fields, rotation curves and rough mass estimates.
The velocity fields obtained appear to be very peculiar, showing non
axi-symmetric distortions and in several cases evidence for multiple
dynamical components.
In this paper we will continue the discussion and  combine the dynamics 
with surface photometric data to model the dynamical and photometric 
masses of the galaxies.  Moreover, morphologies will be discussed 
in detail.

The absolute visual magnitudes of BCGs range from ~$-12$~ to ~$-21$, and 
there is of course no guarantee that our sample of relatively luminous 
BCGs is representative for all BCGs. Nor is there any guarantee that BCGs 
with similar mass or luminosity have similar physical origin. Morphologically 
there exist different types of BCGs (Loose and Thuan \cite{loose:thuan}, 
Salzer et al. \cite{salzer:etal},  Kunth et al. \cite{kunth:1988}, 
Telles et al. \cite{telles:etal}). 
Apart from the varying intrinsic luminosities, the relative intensity 
of star formation varies as well and not all BCGs are starbursts. 
Many BCGs are simply distant dwarf irregulars with  moderately active 
star formation that have been picked up in emission line surveys.
Telles et al. (\cite{telles:etal}) found that BCGs which have irregular 
morphology at faint isophotal levels, are on average more 
luminous than those with regular morphology. Most of the
BCGs in our sample have irregular morphology and thus may have a
physically different evolutionary history as compared to low mass BCGs
with regular morphology.

In 1999 and 2000 we carried out observations of some 15 new BCGs at 
La Silla, extending our sample to lower luminosities, and observing 
time for spectroscopic and photometric followup has been obtained. 
The new sample was selected to extend our luminosity range downwards,
and include galaxies with a minimum H$\beta$ equivalent width 
(sometimes estimated from H$\alpha$) of $\sim 30$ \AA ngstr\"om.
These observations
will be discussed in forthcoming papers. In addition, we obtained
complimentary observations of the H$\alpha$ line widths for the BCGs 
in the current sample, which will be partly used in the present paper.

Throughout the paper we use H$_0 = h_{75} \cdot 75$ km/s/Mpc, with 
$h_{75}=1$. In general, absolute photometric properties like masses,
luminosities and SFR, scale with $h_{75}^{-2}$. Kinematical properties
are not sensitive to the choice of $h_{75}$, but the kinematical
mass estimates scale with $h_{75}^{-1}$ since they involves a length 
scale. Relative properties like mass to light ratios and time-scales
do not depend on  $h_{75}$.

\section{Photometric data and analysis}

In this section we will describe the photometric data available and how 
we use this to obtain realistic estimates of the mass of the stellar
populations in the target galaxies.

\subsection{Photometric data and decomposition}

For most of the galaxies we have extensive optical and near infrared 
images and, in addition, photo-electric photometry (Bergvall and Olofsson 
\cite{bergvall:olofsson}; Bergvall and Olofsson private correspondence).  
These data have been collected at the ESO telescopes on La Silla. The
available photometry for each galaxy is shown in Table \ref{table2}.

The luminosity profiles have been derived by integrating in 
elliptical rings, using the same position angle and inclination as
derived from the kinematical data.  The luminosity profile software
was written in Uppsala and is described in Bergvall and \"Ostlin
(\cite{bergvall:ostlin}), where the photometric properties of several of
the galaxies in the present study is discussed at length.\footnote{Note
that the luminosity profiles used in this paper and those presented in
Bergvall and \"Ostlin (\cite{bergvall:ostlin}) may differ slightly due
to slightly different choices of inclination and position angle for
the ellipses. However, this does not affect the relative contribution
of the two components, and hence not the photometric mass estimates below.}

The luminosity profile of a galaxy traces the spatial distribution of
its stars.  In the ideal case of a uniform stellar population with no extinction, 
the surface brightness directly reveals the stellar mass density. However
if the mass to luminosity ratio of the stellar population varies,
e.g. due to star formation, there will be no simple relation between
the observed luminosity distribution and the real mass distribution of
stars.

\begin{figure}
\resizebox{\hsize}{!}{\includegraphics{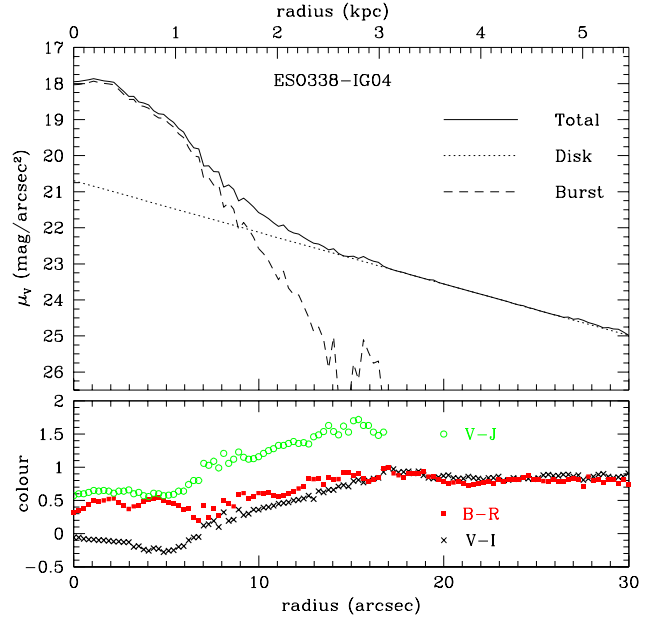}}
\caption[]{Radial luminosity and colour profiles for ESO\,338-IG04.  
The upper panel shows the V-band surface brightness (uncorrected for 
inclination): The total luminosity profile is shown as a solid line,
the fitted disk component is shown as dotted line, and the burst component 
is shown as a dashed line. 
The lower panel shows the radial $V-J, B-R,$ and $V-I$ colour profiles, 
where $B,V$ and $J$ are in the Johnson system, and $R$ and $I$ in the 
Cousins system.   The optical and the near infrared data has been 
obtained at La Silla with NTT+EMMI and 2.2m+IRAC2, respectively 
(see Bergvall and \"Ostlin \cite{bergvall:ostlin}). }
\label{lumprof338}
\end{figure}

In BCGs this problem is serious because, in general, the starburst
population dominates the emission of light, but not the mass. Furthermore,
since the fraction of old stars is a priori unknown (especially in the
starburst region) and the starburst has a rapid luminosity evolution,
the mass to light ratio of the stellar population is very uncertain
and varies with galacto-centric distance. In general, the starburst
region is close to the optical centre of a BCG. In the outskirts $M/L$
may be above one, while in the youngest region of the burst it should
be much lower. To make progress we have taken the following approach:

By looking at the radial colour profiles, we determine the extent of
the starburst.  In the outskirts the colours are  stable (see Fig.
\ref{lumprof338}), indicating that the assumption of a homogeneous
stellar population is plausible here.  Therefore, the luminosity
profile of each galaxy was  decomposed into two
components, a {\em disk} and a {\em burst}. The disk component has
little or no colour gradient and is interpreted as the underlying,
older, population. It has been produced from an exponential fit to the
luminosity profile (usually in the V-band) at radii where the colours
were stable, and, if possible, comparable to the radius of the last
measured point in the rotation curves. In all cases, an exponential law provides a
good fit to the data at these radii. The residual after subtracting
the disk component, we call the burst component. The true photometric
structure of the galaxies at larger radii may be more complicated
(Bergvall and \"Ostlin \cite{bergvall:ostlin}), but this is not of
concern here since we are only interested in estimating the mass of
the stellar population within the extent of the rotation curves.  An
example of the photometric decomposition is shown in
Fig. \ref{lumprof338}. Note how the colours stabilise at the same
radius as the disk takes over from the burst as the main contributor
to the emitted light.  This is the typical situation in the sample
studied here. A representative choice of colour indices for the disk
and the subtracted burst components  are shown in 
Table \ref{table2}. Multi-wavelength colour profiles for all galaxies, 
except ESO~185-13 and Tololo~0341-407, are presented in Bergvall and 
\"Ostlin (\cite{bergvall:ostlin}).

\subsection{Internal extinction}

BCGs in general have low internal extinction (e.g. Mas-Hesse and Kunth 1999),
which probably is a consequence of low chemical abundances and selection
criteria that favour galaxies with blue colours.
The central $E(B-V)$ 
values derived spectroscopic observations of the H$\alpha/$H$\beta$ ratio
and Case B recombination are given in Table \ref{table2}. In some cases, these 
could be overestimated due to underlying Balmer absorption from young A-type
and old stars. Long slit spectra (Bergvall and \"Ostlin \cite{bergvall:ostlin})
indicate that the H$\alpha/$H$\beta$ ratio decreases with radius.  
Moreover, investigations including the UV-continuum suggest that the emission 
lines are more heavily reddened than the stellar
continuum (Calzetti et al. 1994; Mas-Hesse and Kunth 1999). Hence,
applying the $E(B-V)$ values in Table \ref{table2} throughout would overestimate
the internal extinction. We have used the values given in Table \ref{table2}
for the central of the galaxies and assumed that the extinction coefficient 
scales with the thickness of the disk, i.e. the surface brightness. In effect,
the adopted extinction for the burst component is close to the values given in
Table \ref{table2}, whereas the extinction for the disk component is significantly
smaller. In any case, the reddenings in Table \ref{table2} are small and the
uncertainty in the disk $E(B-V)$ values do not seriously affect the derived
photometric masses, as we argue below.

\subsection{Spectral evolutionary synthesis models}

We use a spectral evolutionary synthesis model in combination with 
colour profiles  in the optical and near infrared to estimate the 
mass-to-light ratios of the galaxies. The  model used has been 
developed by Bergvall and is described in Bergvall and R\"onnback 
(\cite{bergvall:ronnback}), \"Ostlin et al. (\cite{ostlin:e338gc}) 
and Bergvall and \"Ostlin (\cite{bergvall:ostlin}). Briefly, it is 
based on stellar evolutionary tracks from the Geneva group (Schaller 
et al. 1992), mostly Kurucz (1992) model atmospheres, and a nebular 
emission component from Cloudy (Ferland 1993). Moreover, pre-main 
sequence stage evolution  from VandenBerg (1985; private correspondence 1986),
and horizontal branch and AGB stages up to the onset of thermal 
pulsations from Castellani et al. (1991), have been included. Empirical 
data are used to extend the evolution to
the tip of the AGB. 
Most of the
galaxies have determined nebular oxygen abundances, which are close to
1/10th of solar (Bergvall and \"Ostlin \cite{bergvall:ostlin}; Masegosa 
et al. \cite{masegosa}).  This value was therefore used for the stellar and
gaseous components in the spectral evolutionary synthesis model.
The used model parameters are summarised 
in Table  \ref{table1}. 

The model also calculates $M/L_{V}$, the V-band mass to light ratio
for the stellar population. This includes the mass of stellar remnants,
and the gas returned to the interstellar medium. Counting only the mass
of stars and remnants would result in slightly lower  $M/L$ values.  
What matters is to what extent the returned gas is used for new star 
formation. Since these galaxies have low metallicity their ISM are not 
heavily polluted and the returned gas will be diluted before any of it 
is taken up in new stars. Hence it is justified to include the returned 
gas in the mass budget for calculating $M/L$.

The models by Bergvall, described in Table 1,  were complimented with 
models produced with the PEGASE.2 code (Fioc and Rocca-Volmerange 1999), 
to check consistency and to further explore the parameter space. In 
particular, models with shorter star formation time-scales (instantaneous 
burst and  e-folding times of 10, 30 and 100 Myr) were used to constrain 
the properties of the central ``burst'' regions.

\begin{table}[tbp]
\caption[]{Spectral evolutionary synthesis model parameters used in $M/L$ determinations.
The IMF slope is defined as $dN/dM \propto M^{-\alpha}$, where $M$
is the stellar mass and $dN$ is the number of stars in the mass
interval $[M, M+dM]$. The classical value by
Salpeter (1955) is  $\alpha = 2.35$.  }
\label{table1}
\begin{flushleft}
\begin{tabular}[t]{ll}
\hline
\noalign{\medskip}
IMF slope $\alpha$ &  2.0, 2.35 and 2.70\\
Lower mass limit $ M_{\rm low}$ & 0.1 and 1.0 \msun \\
Upper mass limit $ M_{\rm up}$ & 30 and 100 \msun \\
Burst duration $\tau$ & 50 and 16000 Myr \\
SFR e-folding time $\beta$ & 0 and 3 Gyr \\
Metallicity & $0.1  Z_{\odot}$ \\
\noalign{\medskip}
\hline
\end{tabular}
\end{flushleft}
\end{table}

\begin{table*}[tbp]
\caption[]{Available photometry and adopted V-band $M/L$ ratios:  }
\label{table2}
\begin{flushleft}
\begin{tabular}[t]{llllllllll}
\hline
\\
Galaxy &  $E(B-V)$	 & Photometry	& Disk$/$Burst colours &  Disk colour&
$M/L_{\rm disk}$ & $M/L_{\rm disk}^{\rm min}$ & $M/L_{\rm disk}^{\rm max}$ & $ M/L_{\rm burst} $\\
 (1) & (2) & (3) & (4) & (5) & (6) & (7) & (8) & (9) \\
\hline
\noalign{\medskip}
\object{ESO 350-38} & $\le$0.25  & $BVRIJHK$ & $B-V = 0.5 / 0.3$ & $V-J = 1.9  $ & 3.0	& 2.0	& 6.0  & 0.05	\\
\object{ESO 480-12} & $\le$0.30  & $BVRIJHK$ & $B-V = 0.3 / 0.0$ & $V-J = 1.6  $ & 2.3	& 1.6	& 3.6  & 0.06	\\
\object{ESO 338-04} & $\le$0.06  & $BVRIJHK$ & $B-V = 0.3/  0.2$ & $V-J = 1.7  $ & 3.0	& 2.1	& 3.8  & 0.05	\\
\object{ESO 338-04B}& $\le$0.06:  & $VIJHK$ & $V-I~ = 0.7 / 0.0$ & $V-J = 1.6  $ 	 & 3.4	& 2.5	& 6.0  & 0.10	\\
\object{ESO 185-13} & $\le$0.16  & $U^aB^aVRI^a$ & $V-R = 0.4 / 0.2$ & 	 & 2.2	& 1.0	& 6.0  & 0.08	\\
\object{ESO 400-43} & $\le$0.22  & $BVRIJHK$ & $B-V = 0.9 / -0.2$ & $V-J = 1.6 $  & 3.1	& 2.2	& 6.5  & 0.06	\\
\object{ESO 400-43B}& $\le$0.17  & $BV^aRI^aJHK$ & $B-R = 0.8 / 0.4$ & $B-J = 2.3 $ 	 & 3.0	& 1.0	& 4.0  & 0.10	\\
\object{Tol 0341-403E} & $\le$ 0.24   & $R$ &   	& & 3.0:	& 1.0:	& 6.0: & 0.03	\\
\object{Tol 0341-403W} & $\le$ 0.29   & $R$ &    & & 3.0:	& 1.0:	& 6.0: & 0.05	\\
\noalign{\medskip}
\hline
\end{tabular}
\end{flushleft}
Description of column contents: (1) Name of galaxy.
(2) The central $E(B-V)$ derived from spectroscopy. The source is Bergvall and \"Ostlin 
(\cite{bergvall:ostlin}), except for ESO~185-13 (Calzetti et al. 1994) and Tololo~0341-403
(Terlevich et al. 1991). For E338-04B no spectroscopic $E(B-V)$ was available and we used
the same value as for ESO 338-04, and flagged it with a colon (:).
(3) Available photometry. Entries with superscript $^a$ indicates
that for this filter only aperture  photometry is available. 
(4) Representative optical colours for the disk and burst regions, corrected for extinction 
(see Sect. 2.2).  $B$ and $V$ are in the Johnson filter system,  $R$ and $I$ in the Kron-Cousins 
filter system. Typical errors are 0.1 or less for the disk, and slightly larger for the burst components.
(5) Representative optical--near infrared colour for the disk components,  errors 0.2-0.3.
(6) shows the best fitting $M/L_V$ values  (in solar units) for the disk component.
(7) and (8) show the 3$\sigma$ upper and lower limits on the disk $M/L_V$, respectively. 
The uncertain $M/L_V$ values for Tololo 0341-403 are indicated with colon, see text for
further explanation.
(9) Best fitting $M/L_V$ estimate for the burst component.
\end{table*}

\subsection{$M/L$ values}

The ensemble of model predictions characterised in Table \ref{table1}  were 
compared to the observed broad-band colours of the galaxies. The colours of 
the disk and burst components where modelled separately. 
Taking the observational uncertainties 
into account, we determine which models reproduce the disk colours 
within $\sim 3 \sigma$ and read off their $M/L$ value. 
Some models were discarded because they give unreasonable $M/L$ values at 
certain ages, e.g. short burst 
models with lower mass limit $M_{\rm low} = 1 M_\odot$ and ages larger than 
a few Gyr. Such models are in any case not very realistic.
Apart from this, the best-fitting $M/L$ is generally well defined, as is
also the minimum allowed $M/L$ value.  The upper limit to the $M/L$ value
is generally less well constrained.

In many cases there is a ``good degeneracy'', in the sense that while
the best fitting age is sensitive to the choice of model parameters,
the corresponding $M/L$ values are very similar. We illustrate this
with the following example: Compared to a standard $\alpha = 2.35$
model, a model with a steeper $\alpha = 2.85$ IMF, will for each
time-step have redder colours and higher $M/L$.  Fitting a galaxy's
colour to this steep IMF model will produce a lower age, and this will
to first order compensate for the change in $M/L$. The same effect occurs
if one varies the metallicity. Similarly, increasing the 
assumed $E(B-V)$ value decreases the modelled age and $M/L$ but this 
effect is to a first order compensated by the increasing amount
of absorbed emission to correct for.

In this paper, we have only considered single power-law IMFs. The use 
of broken power-law IMFs with two or more segments (e.g. Miller \& Scalo 1982,  
Kroupa et al. 1993, Scalo 1986, 1998) could decrease the modelled $M/L$ 
values somewhat, up to a factor of 2 at large ages.  Hence, if the IMF 
in our galaxies has a flatter low mass range, the real $M/L$ values would 
approach the lower limits given in Table 2. On the other hand, the 
possible extension of the IMF to objects with sub-stellar masses would 
increase   $M/L$ by $\sim$ 10\% for an IMF with relatively flat slope 
in the regime below 1 $M_\odot$.

In general, models with $M_{\rm low} = 0.1$ \msun, $M_{\rm up} = 100$
\msun, $\alpha = 2.35$ and a continuous but exponentially decaying
($\beta = 3$ Gyr) star formation rate give the best fit to the
observed colours of the disk component. The best fitting ages are
found between 2.5 and 8 Gyr for the different galaxies (between 1 and
16 Gyr considering the lower and upper $3 \sigma$ limits).  Of course
there might be even older components present, that however cannot make
a significant contribution to the mass within the modelled radii.  The
age of the halo stellar populations is further discussed in Bergvall
and \"Ostlin (\cite{bergvall:ostlin}).  The resulting $M/L$ values for
the disk component are shown in Table \ref{table2}. For ESO\,185-13, we
have only optical surface photometry in $V$ and $R$ and no near-IR data
and consequently its $M/L_{\rm disk}$ values are less certain.
For one galaxy
(\object{Tololo 0341-407}), we did not have multicolour data, and instead we used
the median value $M/L_{\rm disk}$ for the other galaxies, and the extreme of
$M/L_{\rm disk}^{min}$ and $M/L_{\rm disk}^{max}$ for the allowed range;
these values  are flagged with a colon (:) in Table  \ref{table2}.

It was more difficult to constrain the $M/L$ values for the burst
components, since there is a large dependence on model parameters and
strong colour gradients.  Moreover, ionised gas emission is important
and may be non-local, i.e. the emitting gas is displaced with respect
to the ionising source and e.g. found in filaments (\"Ostlin et al.
1998). The contribution of line and continous emission from ionised gas
has the effect of making the burst colours presented in Table \ref{table2}
redder. Part of the difficulty may arise because the burst population
is not completely coeval, but includes a mix of stars with ages
differing up to several 10 Myrs. For the burst, we used also the observed 
H$\alpha$ equivalent-widths to constrain its age and $M/L$. We were able to
constrain that in all cases $0.01 < M/L < 0.2$, except when using a very
steep IMF ($\alpha \ge 3.0$), which however gives worse fits to the data.  
We have used the values given in Table 2  as our estimates for the burst,
with a generous allowed interval of $M/L=0.01$ to $M/L=0.4$, common for
all galaxies.  Within
this range in $M/L$, the burst always make a minor contribution to the
total photometric mass.

\begin{table*}[tbp]
\caption[]{Absolute magnitudes, and mass estimates of disk and burst components. 
 }
\label{table3}
\begin{flushleft}
\begin{tabular}[t]{llllllllllll}
\hline
Galaxy & R & Disk & Burst & ${ M}_{\rm disk}$ & ${ M}_{\rm
burst}$ & ${ M}_{\rm ph}$ & ${ M}_{\rm rot}$ & Note\\

& kpc   &  $M_V$  &  $M_V$ & ${10^9  M_{\odot}}$  & 
 ${10^9  M_{\odot}}$  & ${10^9  M_{\odot}}$ & ${10^9  M_{\odot}}$    \\

(1) & (2) & (3) & (4) & (5) & (6) & (7) & (8) & (9) \\

\hline
& & & \\
{ESO 350-IG38}  &  6.0 & -19.3  & -20.4   & 13$_{-4.4}^{+13}$ 
& 0.64$_{-0.51}^{+4.5}$ & 13.9$_{-4.9}^{+18}$ & 0.87$_{-0.60}^{+2.1}$ & Rmax, non decomp.\\
\\
~~~~~~~~~~``~~~ & 0.4 &  -15.8 & -15.4  & 0.51$_{-0.17}^{+0.51}$ 
& 0.006$_{-0.005}^{+0.050}$ & 0.52$_{-0.18}^{+0.56}$ & 0.60$_{-0.40}^{+0.90}$  & Vmax, non decomp. \\
\\
~~~~~~~~~~``~~~ &  5.2 & -19.2  & -20.4   & 13$_{-4.2}^{+13}$ 
& 0.64$_{-0.52}^{+4.5}$ & 13$_{-4.7}^{+17}$ & 1.6$_{-1.1}^{+3.2}$ & DVF: 1st comp.\\
\\
~~~~~~~~~~``~~~ & 1.0 &  -17.3 & -18.1 & 2.1$_{-0.7}^{+2.1}$ 
& 0.078$_{-0.062}^{+0.54}$ & 2.1$_{-0.75}^{+2.6}$ & 1.3$_{-0.90}^{+5.6}$  & DVF: 2nd comp.\\
\\
{ESO 480-IG12}   & 5.9 & -18.4 & -19.2 &  4.6$_{-1.4}^{+2.6}$ 
& 0.25$_{-0.21}^{+1.4}$ & 4.8$_{-1.6}^{+4.0}$ & 17$_{-8.2}^{+17}$ & Rmax\\
\\
{ESO 338-IG04}   & 2.5 & -17.3 & -19.0 & 2.2$_{-0.65}^{+0.58}$  
& 0.17$_{-0.14}^{+1.2}$ &  2.3$_{-0.80}^{+1.8}$  & 0.57$_{-0.50}^{+1.8}$ & masked Rmax \\
\\
~~~~~~~~~~``~~~         & 1.9  & -17.0 & -19.0 & 1.6$_{-0.5}^{+0.4}$  
& 0.17$_{-0.14}^{+1.2}$ &  1.8$_{-0.62}^{+1.6}$  & 0.86$_{-0.70}^{+2.0}$ & masked Vmax \\
\\
{ESO 338-IG04B}   &  3.6 & -17.7  & -16.4 & 2.8$_{-0.75}^{+2.2}$  
& 0.032$_{-0.03}^{+0.10}$ & 2.9$_{-0.78}^{+2.3}$ & 5.7$_{-2.9}^{+6.0}$ & Rmax \\
\\
{ESO 185-IG13}   &  3.8  & -18.7  & -18.7 & 5.6$_{-3.1}^{+9.7}$
  & 0.21$_{-0.18}^{+0.85}$ & 5.8$_{-3.3}^{+11}$ & 1.5$_{-0.90}^{+1.9}$ & Rmax \\
\\
{ESO 400-G43}    &  10.9 & -19.2  & -20.0  &  12$_{-3.4}^{+13}$  
& 0.52$_{-0.43}^{+2.9}$ & 12$_{-3.9}^{+16}$ & 0.29$_{-0.20}^{+0.70}$ & Rmax \\
\\
~~~~~~~~~~``~~~   &   5.7 & -18.9  & -20.0 &  9.9$_{-2.9}^{+11}$  
& 0.50$_{-0.42}^{+2.9}$ & 10$_{-3.3}^{+14}$ & 0.66$_{-0.50}^{+1.1}$ &  Both \\
\\
~~~~~~~~~~``~~~    &  1.1 & -16.9  & -18.2  &  1.6$_{-0.46}^{+1.7}$  
& 0.094$_{-0.078}^{+0.53}$ & 1.7$_{-0.54}^{+2.3}$ & 0.60$_{-0.30}^{+0.70}$  & Vmax  \\
\\
{ESO 400-G43B}   & 4.8 &  -17.9 & -18.1 &  3.6$_{-2.4}^{+1.2}$  
& 0.15$_{-0.14}^{+0.46}$ & 3.8$_{-2.6}^{+1.7}$ & 3.3$_{-1.5}^{+2.1}$ & Rmax \\ 
\\
{Tololo 0341-403E}   & 1.6  &  -14.6 & -15.9  & 0.17$_{-0.12}^{+0.17}$ 
 & 0.006$_{-0.004}^{+0.073}$ & 0.18$_{-0.12}^{+0.25}$ & 0.17$_{-0.1}^{+0.5}$  & Both \\
\\
{Tololo 0341-403W}   & 1.9  & -15.6 &  -16.5 &  0.43$_{-0.29}^{+0.43}$  
& 0.018$_{- 0.014}^{+0.12}$ & 0.45$_{-0.30}^{+0.56}$ & 0.26$_{-0.20}^{+0.50}$ &  Both \\
\noalign{\medskip}
\noalign{\medskip}
\hline
\end{tabular}
\end{flushleft}
Description of column contents: (1) Name of galaxy
(2) Radius (in kpc) within which the properties in columns 3 to 8 have been evaluated. 
(3) Absolute V-band magnitude of disk component. 
(4) Absolute V-band magnitude of burst component. 
(5) Estimated photometric mass of the disk component. 
(6) Estimated photometric mass of the burst component.
(7) Total photometric mass of the disk+burst component. 
(8) Rotational mass. The quoted uncertainties reflect the uncertainty in: the
 parameter $f$, the inclination and the rotational velocity, see Paper I. 
(9) Comments: For some galaxies we present mass estimates at more than one radii,
corresponding to certain features in the rotation curve (see Paper I).
``Rmax'' and ``Vmax'' means that the mass was evaluated at the last measured 
point in the rotation curve, and at the radius of maximum velocity, respectively.
``Both'' means that the mass was evaluated at the last point where both
the approaching and receding sides could be measured.
For ESO\,350-38 we present mass estimates obtained both from the model
decomposing the velocity field into two dynamical components (DVF) and 
for the ``raw'' velocity field. For ESO\,338-04 some points in the velocity 
field had to be masked in order to at all be able to derive a rotation curve. 
NB: For all entries, the quoted uncertainties represent the maximum expected 
deviation, i.e. rather 3-sigma than 1-sigma deviations.  
\end{table*}

\begin{figure}
\resizebox{\hsize}{!}{\includegraphics{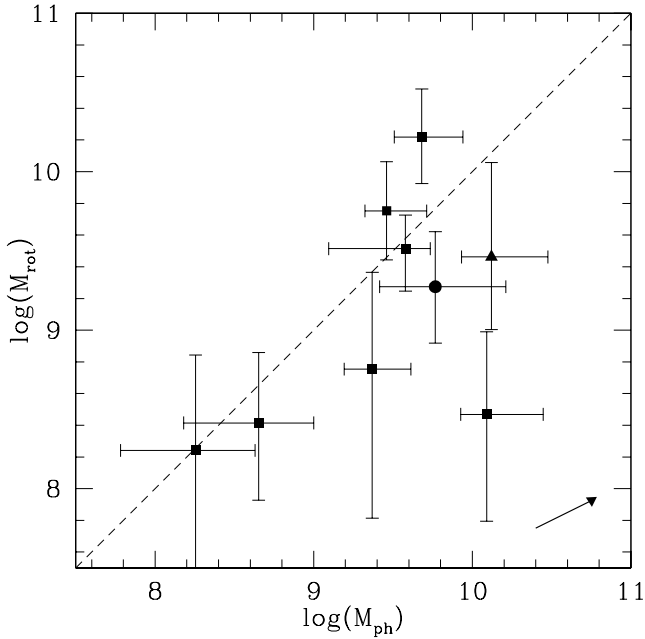}}
\hfill
\caption[]{Comparison of the photometric and dynamical (rotational)
masses. The X-axis shows the logarithm (base 10) of the derived
photometric mass in units of $M_{\odot}$.  The Y-axis shows the 
logarithm of the rotational mass.  The data-points are taken from
Table 3, except  for ESO\,350-38 (filled triangle) and ESO\,185-13 
(filled circle), where $M_{\rm rot}$  is the sum of the rotational 
masses of the two distinct components in the velocity field (see Paper I).  
The diagonal dashed line shows the location of galaxies for which  
the photometric and rotational masses are equal.
The arrow in the lower right shows the effect of going from 
H$_0 = 75$ to 50 km/s/Mpc.
 } 
\label{comparison}
\end{figure}

\begin{table*}[tbp]
\caption[]{Derived global properties. 

}
\label{table4}
\begin{flushleft}
\begin{tabular}[t]{lrrccrrrrr}
\hline
Galaxy & $SFR$ & $M_{\rm ph,26}$ & $M/L_{26}$ & $SFR/M_{\rm ph,26}$ &
 $\tau_{\rm ph}$ & $\tau_{\rm b}$ & $r_{\rm e,d}$ &   $\sigma$ & $M_\sigma$ \\
       & $M_{\odot}$/yr	& $10^{9} M_{\odot}$	& 		& 10$^{-9}$ yr$^{-1}$ 	
& Gyr & Myr & kpc  &km/s & $10^{9} M_{\odot}$ \\
(1) & (2) & (3) & (4) & (5) & (6) & (7) & (8) & (9) & (10) \\
\hline
\noalign{\medskip}
{ESO 350-IG38} 	 & 18.5	& 15.5	& 0.87  & 1.19 	& 0.84 	& 34	& 2.6	& 81	& 18.8	 \\
{ESO 480-IG12} 	 & 4.1	& 5.3	& 0.83 	& 0.77 	& 1.29	& 60	& 2.6 	& 51	& 7.7	 \\
{ESO 338-IG04}	 & 3.2	& 4.0	& 0.85 	& 0.80  & 1.24	& 53	& 2.2 	& 43	& 4.6	 \\
{ESO 338-IG04B}	 & 0.3	& 3.2	& 2.53	& 0.08  & 12.1	& 120	& 1.7 	& 29	& 1.6	 \\
{ESO 185-IG13} 	 & 5.4	& 7.1	& 1.21 	& 0.76  & 1.32	& 40	& 2.0 	& 40	& 3.7	 \\ 
{ESO 400-G43}	 & 11.3	& 12.7	& 1.07 	& 0.89  & 1.12	& 45	& 3.1 	& 49	& 8.3	 \\
{ESO 400-G43B}	 & 1.3	& 4.9	& 1.52 	& 0.27  & 3.70	& 110	& 2.6 	& 34	& 3.4 	 \\
{Tololo 0341-403E}	 & 1.3	& 0.3	& 0.98 	& 4.38  & 0.23	& 5	& 1.6 	& 34	& 2.1	 \\ 
{Tololo 0341-403W}	 & 0.9	& 1.0	& 1.43  & 0.90  & 1.11	& 19	& 2.2 	& 36	& 3.3	 \\
\noalign{\medskip}
\hline
\end{tabular}
\end{flushleft}
Description of column contents: (1) Name of galaxy. (2) Star formation rate (SFR) 
derived from the integrated H$\alpha$ luminosity, uncorrected for internal extinction.
 (3) Integrated (disk+burst) 
photometric mass within the $\mu_{V} = 26 $ mag/arcsec$^2$ isophote. (4) Integrated V-band mass 
to light ratio within $\mu_{V} = 26 $ mag/arcsec$^2$. (5) SFR per unit (photometric) 
mass  of the galaxy.  (6) Inverse of $SFR/M_{\rm ph,26}$, giving the time-scale 
for building up $M_{\rm ph,26}$  with the current SFR. (7) Time scale for accumulating
the observed burst mass with current SFR. (8) Effective (half-light) radius of the 
disk component. (9)  Integrated H$\alpha$ line-width. (10) Mass inferred from the 
H$\alpha$ line-width.
\end{table*} 

\subsection{Star formation rates and time-scales}

With the aid of the spectral synthesis code described in Sect. 
2.2., the integrated H$\alpha$ luminosities have been used to derive 
the total star formation rates in the observed galaxies. For a Salpeter
IMF with mass range $0.1-100 M_\odot$, and  with $L(H\alpha)$  in Watts,
 the SFR expressed in $M_{\odot}/{\rm yr}$ is:

  $$SFR =L(H\alpha)/1.51\cdot 10^{34}$$

The integrated SFR for all galaxies are given in  Table \ref{table4}; 
the H$\alpha$ luminosities are given in Table 5 in Paper I. Note
that these are lower limits since the  H$\alpha$ luminosities were
derived  assuming no extinction. In addition, some of the very youngest 
star forming regions may be completely embedded.
The central $E(B-V)$ estimates, derived from from spectroscopic
H$\alpha$/H$\beta$ values, are in the range $E(B-V)=0 - 0.3$, corresponding
to  correction factors of $1.0$  to  $2.0$ for  $L(H\alpha)$. However, slit
spectra indicate that $H\alpha/H\beta$ decrease outside the centre, and since 
we do not know the ``global'' $E(B-V)$ values we did not apply any correction 
to the values in Table \ref{table4}. The derived star formation rates span 1 
to 20 $M_{\odot}$ per year for the BCGs, whereas the  companion to ESO\,338-IG04
has a lower SFR. 

Together with the photometric and rotational mass estimates 
the integrated SFR estimates can be used to derive mass
averaged values. In Table \ref{table4} we give the SFR normalised 
to the photometric galaxy masses. The normalisation using rotational 
mass would in some cases (when there is  an apparent 
dynamical mass deficiency) overestimate the mass averaged SFR, and
this quantity is not presented explicitly.

The inverse of the mass averaged  SFRs also provide $\tau_{\rm ph}$, the time 
scale for building up the observed photometric mass in stars with the 
current SFR. In general, $\tau_{\rm ph}$ is close to 1 Gyr, meaning that the
current SFR is one order of magnitude larger than the average past SFR. 
Thus, these galaxies are true star bursts in the sense that their SFR 
is an order of magnitude higher than what is sustainable over a Hubble 
time. Note that the modelling in Sect. 2.3  indicate ages of several Gyrs 
for the underlying populations. For the two companions we find larger
values of $\tau_{\rm ph}$. In the eastern component of Tololo\,0341-407 
$\tau_{\rm ph} \sim 0.25$ Gyr, indicating a momentary very enhanced SFR,
although its photometric mass is quite uncertain.

Similarly, we can define $\tau_b = M_{burst}/SFR$, the time scale for
building up the observed burst mass. Hence $\tau_b$ gives an order of 
magnitude estimate of how long the present burst has been active. Values
of  $\tau_b$ are given in Table 4. If the SFR has passed its peak, 
$\tau_b$ will overestimate the burst duration. In ESO\,338-IG04 we find
$\tau_b \sim 50$ Myr, in good agreement with the burst duration derived 
from the age distribution of young star clusters resolved with the HST 
(\"Ostlin et al. 1998).

Since we corrected for extinction in deriving $M_{ph,26}$ and $M_{burst}$,
but not when deriving the SFR, the timescales $\tau_{\rm ph}$ and 
$\tau_{\rm b}$  may be slightly overestimated, by up to a factor of two.

Another time scale can be constructed for those galaxies where 
the H{\sc i} content is known, i.e. the gas consumption time scale. 
All galaxies in the sample have determined H{\sc i}-masses or upper limits.
The gas consumption time scale ranges from the extreme value of 5 Myr 
(sic) for ESO\,350-IG38 (which has an upper limit of $M_{\rm HI} 
< 10^8$ \msun, but see Sect. 4.1) to $\sim 3$ Gyr for the companion of 
ESO\,400-G43. All BCGs have gas consumption time-scales smaller than 1 Gyr,
 showing that the present SFRs are unsustainable.

\subsection{Photometric mass estimates, and comparison with rotational masses}

Photometric masses were derived by integrating the luminosity
profiles for the disk and burst components and using their
corresponding $M/L$ values.  The photometric masses were evaluated at
radii taken from characteristics in the rotation curves (e.g. the last
measured point or the radius of maximum velocity) to enable comparison with
the kinematical mass estimates.  In general, the luminosity profiles were
integrated out to a radius corresponding to the last point in the
rotation curve.  In Table \ref{table3}, we provide the photometric
mass estimates, with their lower and upper limits.  We also provide
the dynamical mass estimate from Paper I, with the lower and upper
limits (taking into account the uncertainty in inclination, the
intrinsic dispersion, the uncertainty in the parameter $f$ and the
difference between different rotation curves or decomposition-models for the
individual galaxies). In general, estimating the mass from the 
rotation curve will underestimate the dynamical mass since most likely
these galaxies are not pure rotators. Random  motions of the gas could 
provide significant dynamical support for   our BCGs as indicated by 
their line widths (Table 4).
For rotation curves with low amplitude and for the central regions, 
the contributions from velocity dispersion  may dominate over the 
rotational component (however, see Sect 2.7).

For some galaxies there are several entries in
Table \ref{table3}.  This may be the case e.g. when the rotation curve is not
monotonously rising, or when there are several dynamical components. 
In those cases the rotational masses are derived for the relevant
radius in question and the photometric mass is given for the same
radius.  For instance, for ESO\,350-IG38 we present mass estimates 
for the last point in the rotation curve, for the maximum
rotational velocity and for the two components in the decomposed
velocity field (see Paper I). In Fig. 2, the photometric and 
rotational mass estimates are compared (for reasons of clarity 
we plot only one point for each galaxy). 

In all cases, except for one of the companions, the burst component
dominates the optical luminosity (see Table 3). However, the integrated 
burst mass only makes up 2 to 5\% of the total photometric mass. In 
Table 4, we give $M_{\rm ph,26}$,
the photometric masses integrated out to the $\mu_V = 26 $ mag/arcsec$^2$
isophote. This should be close to the total photometric masses, as
the disk component outside these radii should contribute only on the
order of one percent of the total mass if the $M/L$ values are similar.
In Table 4, we also present $M/L_{26}$, the global mass to light ratio, 
derived from $M_{\rm ph,26}$ and the integrated disk+burst light within 
the corresponding radius.

From Table 3 and Fig. 2 one sees that ESO\,480-IG12 and ESO\,338-IG04B  
are examples in which dark matter is needed within the extent of 
the rotation
curves. In ESO\,400-IG43B, Tololo\,0341-403E and Tololo\,0341-403W  the 
photometric and rotational masses are in good agreement.

ESO\,350-IG38, ESO\,338-IG04, ESO\,185-IG13 and ESO\,400-IG43 present a deficit in 
the rotational mass estimates, as compared to the photometric mass.   
This result can be
understood if the systems are not primarily rotationally supported or
if the rotation curve does not reflect potential motions, either because
the dynamics are not relaxed or if the H$\alpha$ emitting gas is not
moving like the stars, even on a large scale.  The latter is not so likely
although the H$\alpha$ velocities are probably also affected by winds
and expanding bubbles.  To settle this definitely, the stellar velocity
field must be compared with that of the ionised gas. The case of
ESO\,338-IG04 is a clear example of a non-relaxed system in which such a
discrepancy between the two mass estimates does exist.

The photometric mass scales with $h_{75}^{-2}$, whereas the rotational 
mass ($M_{\rm rot}$) scales as $h_{75}^{-1}$. Hence, the ``mass 
deficiency'' apparent in some galaxies gets worse with lowering of the Hubble 
constant from the adopted value of 75 km/s/Mpc.

\subsection{Do the H$\alpha$ line widths trace mass?}

Terlevich and Melnick (1981) showed that giant H{\sc ii}-regions display a 
positive correlation between radius, the H$\beta$ luminosity and the 
H$\beta$ emission line width.  They have argued that this reflects an 
underlying mass relation and that the H$\beta$ line width traces virial 
motions.  This is somewhat surprising since one expects also other 
processes, such as SNe feedback, to contribute to the emission line widths.
H{\sc ii} galaxies (galaxies with H{\sc ii} region like spectra,i.e. 
$\sim$BCGs) seems to follow the same relation, suggesting that they are 
scaled up versions of giant H{\sc ii}-regions (Melnick et al. 1987). This 
also makes it possible to use H{\sc ii} galaxies as standard 
candles to probe cosmological parameters like the H$_0$ and  $\Lambda$
(Melnick et al. 1988, 2000).  

\begin{figure*}
\resizebox{\hsize}{!}{\includegraphics{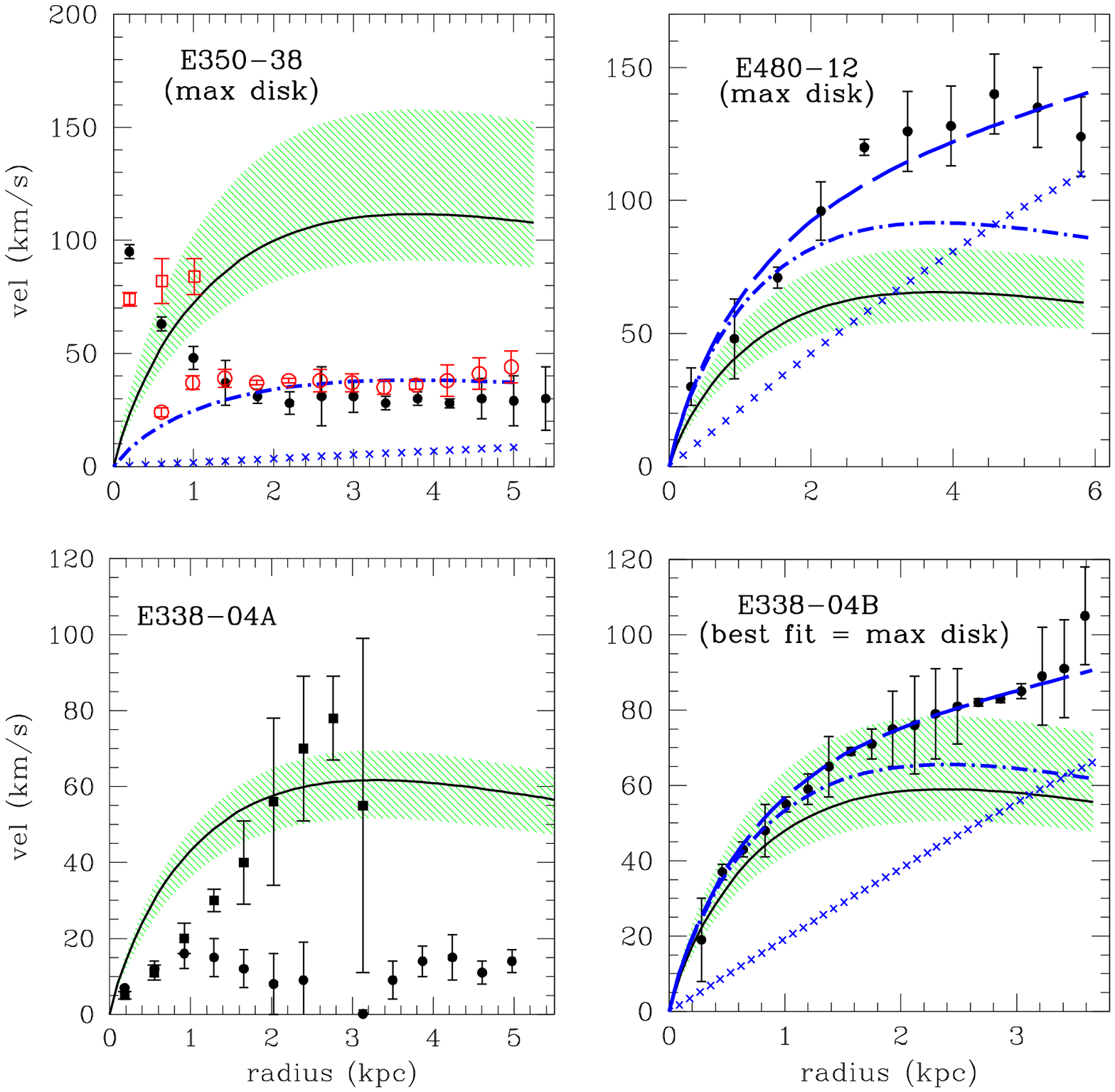}}
\caption[]{Mass models for ESO\,350-IG38, ESO\,480-IG12, ESO\,338-IG04 and
ESO\,338-IG04B. The filled circles with error bars show the observed
rotation curves (see Paper I).  The solid line is the photometric rotation curve 
for the disk component; and  the  shaded area is the allowed range 
based on the uncertainty in $M/L$.  The dash-dot line is the
stellar disk component from the mass model and the crosses are the halo
component; the total model (disk+halo) is shown by the thick long dashed
line (only for those cases where the halo component is non-zero).
For  ESO\,350-IG38, the rotation curve resulting from the non-decomposed 
velocity field is shown as filled circles. The main component of the rotation 
curve for the decomposed velocity field is shown as open circles, and the
secondary counter-rotating component is shown as open squares. The dynamical
mass model was fitted to the main component of the decomposed velocity field.
For ESO\,338-IG04,  the receding side of the rotation curve is shown
as filled circles and the approaching side as filled squares. No disk
and halo component are shown since it was impossible to construct a dynamical 
mass model. }
\label{massmod1st4}
\end{figure*}

If the  Balmer emission line widths indeed reflect virial motions, they 
may be used to derive the dynamical mass of a galaxy. In Table 4, we 
provide, $r_{e,d}$, the effective radii of the disk components and $\sigma_{{\rm H}\alpha}$, 
the measured  H$\alpha$ line-widths ($\sigma = FWHM/2.35$). The 
$\sigma_{{\rm H}\alpha}$ values are the average line-widths where 
each pixel is weighted by its H$\alpha$ intensity. 
They were derived from the velocity dispersion maps\footnote{Except for 
ESO\,338-04, the $\sigma_{{\rm H}\alpha}$ values given in Table 4 were 
derived from  complimentary Fabry-Perot observations utilising an
interferometer with large spectral range and were obtained at
ESO/La Silla in September 1999 and 2000. For ESO\,338-04 $\sigma_{{\rm H}\alpha}$
is derived from the data presented in Fig. 7 in Paper I.} rather than from
the integrated profiles. Hence the line widths given in Table 4 are not 
influenced by the overall rotation of the galaxies. The mass which can be 
supported by velocity dispersion, $M_\sigma$,
was estimated from $M_\sigma = 3 \cdot 1.6  \cdot r_{e,d}  \cdot \sigma^2 / G$, 
(Guzm\'an et al. 1996, Bender et al. 1992). With  $M_\sigma$  in $M_\odot$, 
$r_{e,d}$ in kpc, and $\sigma$ in km/s, this becomes:
 
$$M_\sigma  = 1.1\cdot10^6  \cdot r_{e,d}  \cdot \sigma^2  $$

We caution 
that this is a quite crude estimate, which assumes that the line width
is due to virial motions in response to the gravitational potential of
the underlying disk component. If the bursts are caused by infalling 
material, which is not virialised, this is not necessarily true.
Since $r_{e,d}$ scales linearly with the inverse of $h_{75}$, so $M_\sigma$ 
does as well.

The interesting point is that, for the galaxies which present a 
deficit in the rotational mass estimate, we can solve the mass
discrepancy by invoking that these systems are suported by velocity 
dispersion instead of rotation. This can bee seen by comparing   $M_\sigma$
and  $M_{\rm ph,26}$ in Table 4. However, this does not solve the problem 
of why some of our galaxies have such strangely shaped  rotation curves,
e.g. rapidly declining, as for ESO\,350-IG38 and ESO\,400-G43. In general, the
mass discrepancies occur outside the centre, which dominates the measured
line widths. The ultimate test of the importance of the underlying gravitational 
potential on the H$\alpha$ line widths would be to derive the stellar velocity 
dispersion.  
Kobulnicky and Gebhardt (2000) showed that the central velocity dispersion 
derived from stars and ionised gas in general agree in a sample of late type 
galaxies.

\begin{figure}
\resizebox{\hsize}{!}{\includegraphics{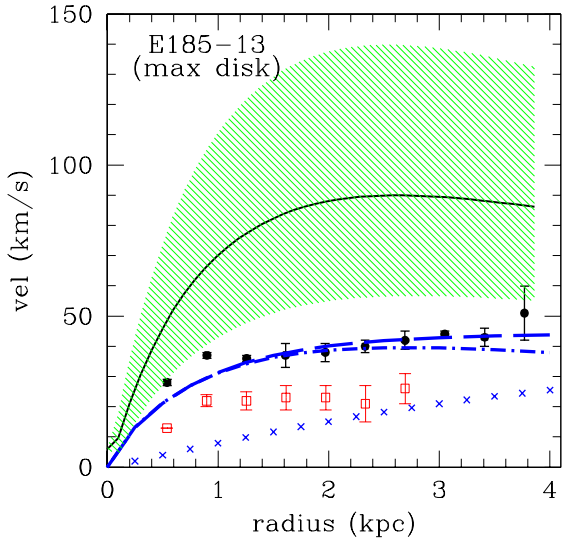}}
\caption[]{Mass model for ESO\,185-IG13 employing the ``maximum disk'' method.
The main component of the rotation curve from the decomposed velocity field 
is shown as filled circles, and the secondary counter-rotating component is 
shown as open squares. The dynamical disk and halo components were fitted
to the main component.
For further explanations see caption of Fig. \ref{massmod1st4}.}
\label{massmod185}
\end{figure}

\begin{figure*}
\resizebox{\hsize}{!}{\includegraphics{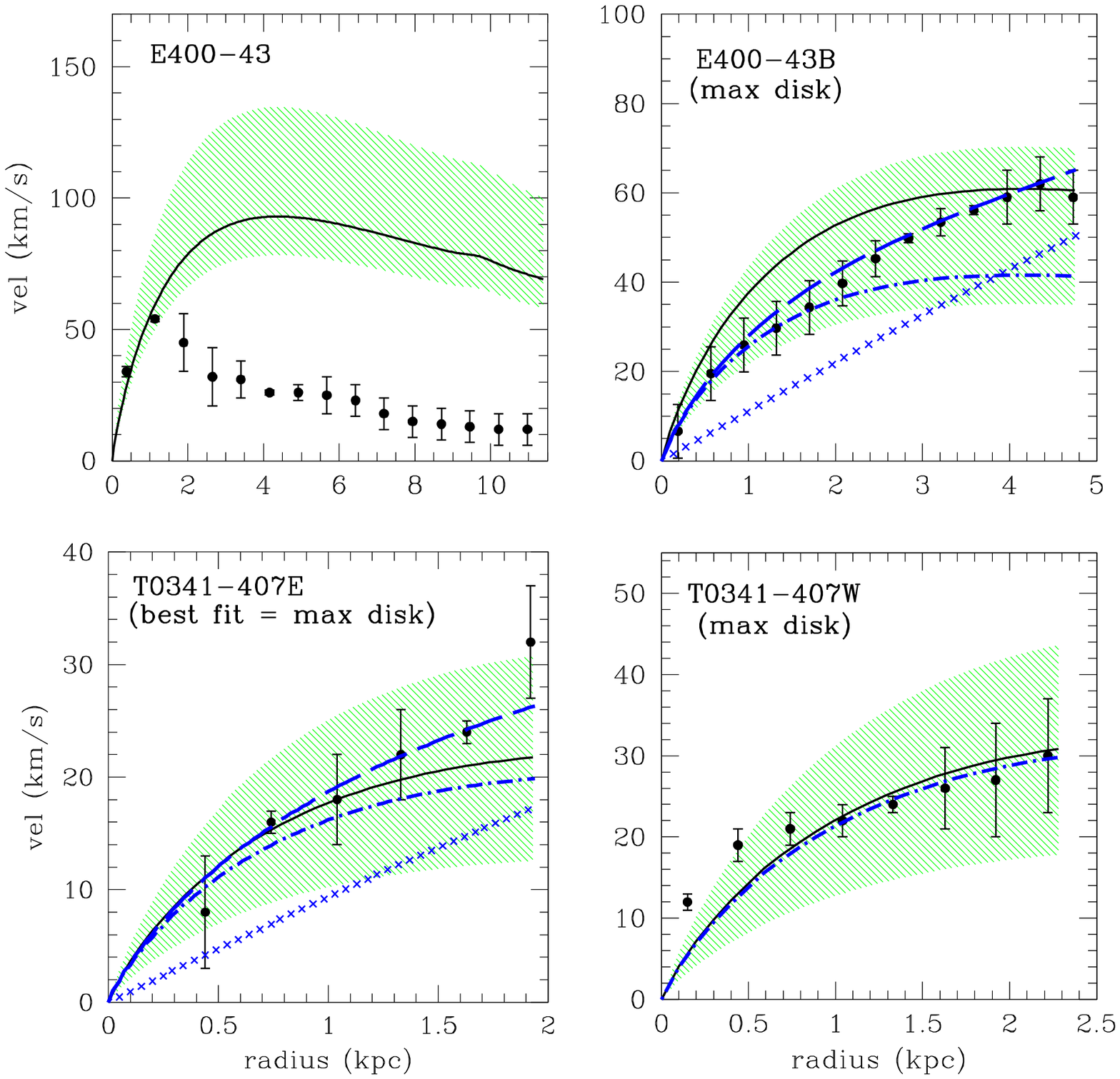}}
\caption[]{Mass model for ESO\,400-G43, ESO\,400-G43B, Tololo\,0341-407E and
Tololo\,0341-407W.  For further explanations see caption of Fig. \ref{massmod1st4}.
For ESO\,400-G43  no disk and halo component are shown since it was impossible 
to construct a dynamical  mass model. }
\label{massmodlast4}
\end{figure*}

\section{Dynamical mass models}

Above and in  Paper I, we used the rotation curves to construct simple 
integrated dynamical mass estimates, based on the formula given by Lequeux (1983). 
In this section we will briefly discuss more detailed mass models involving
a dark matter halo, and which also take  the radial behaviour of the luminosity
profile and the rotation curve into account. 

Given the luminosity profiles for the burst and disk components, and their 
associated $M/L$ values (see Table 2), its is possible to predict what 
the rotation curve should look like, under the assumption that  there is no 
dark matter. We call these {\it photometric rotation curves}. 
These photometric rotation curves are  
the minimum allowed velocities for rotational support. In Figs. 3 to 5, these
are shown as full drawn lines with a hatched region for the allowed interval
(given by the $M/L$ intervals in Table 2).
We show only the disk component, since the burst component in all cases gives
a neglible contribution.  

Dynamical mass models, based on the observed rotation curve and the  
observed shape of the disk and burst luminosity profiles, were also constructed.
Again, the burst component gives a neglible contribution and for reasons 
of clarity we do not include it in the presented mass models.

The technique used to construct the mass models is described in Carignan \& 
Freeman (1985) and the code we used was kindly provided to us by Claude Carignan.  
Both "maximum disk" and "best fit" approaches were undertaken.  A $\chi^2$ minimization
technique is used in the three-parameter space of the model.  These
parameters are: The mass-to-light ratio of the stellar 
disc ($M/L_{disk}^\star$), the core radius ($r_{\rm halo}$) and the one-dimensional velocity 
dispersion ($\sigma_{\rm halo}$) of the dark isothermal halo.  
The mass models are presented in Figs. 3 to 5: the dash-dot lines gives the 
disk component, the crosses give the halo component and the long dashed lines 
give the total disk+halo component.
  The latter should be compared with the observed rotation curve which is shown
as dots with errorbars (here the error bars reflect the discrepancy between   
the approaching and receeding sides). In general we show the maximum disk models 
since they provide  upper limits to the mass contained in stars.

When viewing Figs 3 to 5, the galaxies fall in two categories: Four of the
galaxies (ESO\,350-38, ESO\,400-43, ESO\,185-13 and ESO\,338-04) have observed rotation 
curves that are well below the predicted photometric rotation curves.
This was expected in view of the mass dicrepancies discussed in Sect. 2.5. In 
ESO\,350-38 and ESO\,185-13 the shape of the rotation curve and the disk component 
of the mass model agree reasonably well when using the primary component of the
decomposed velocity fields, but the observed rotation is to slow (marginally in ESO\,185-13,
significantly in ESO\,350-38). ESO\,185-13 could possibly be saved by adopting
an IMF with a flat low mass part, if the dynamical mass has also been underestimated
(e.g. by overestimating the inclination), but not ESO\,350-38.  For ESO\,400-43 and 
ESO\,338-04 it was even impossible to construct a mass model, due to the divergent 
shapes of the observed and photometrical rotation curves. 

In the other galaxies, there is no apparent deficiency of rotation. In ESO\,480-12
and ESO\,338-04B we see a  significant contribution from a dark matter halo to the
dynamics. In ESO\,400-43B and Tololo\,0341-407E there is marginal evidence for dark
matter: the fit is improved by including a dark halo, but the luminous matter 
still dominates within the extent of the rotation curve. In Tololo\,0341-407W we 
see no signature of a dark halo. In these 5 cases the disk components agree
rather well with the photometric rotation curves.

The galaxy ESO\,338-IG04B has both well behaved dynamics and photometric 
structure.  For this galaxy, the maximum disk and best fit models give the 
same result: $M/L_{disk}^\star=4$, core radius of the dark halo $r_{halo}=9$ 
kpc, and a halo velocity dispersion $\sigma_{halo}=100$ km/s.  The 
99.9\%  confidence interval  is $M/L_{disk}^\star = 3.2$ to $4.8$. The 
photometric estimate is $M/L_{disk}=3.4$  with a 3 sigma confidence interval 
$2.5$ to $6.0$. A photometric mass   slightly lower than the dynamical
one is expected with the presence of an H{\sc i} disk. Thus there is a very 
good agreement between the purely dynamical and the purely photometric $M/L$ 
estimates, which give us confidence in our 
photometric $M/L$ estimating procedure.

\section{Discussion  of individual objects}

In this section, we discuss the kinematics and morphology of the
individual targets.  In addition, the velocity fields are thoroughly
described in Paper I.  For all galaxies we show isovelocity contours
for the H$\alpha$ emitting gas, overlaid on broad band images. The 
labels of the isovelocity contours are given in Paper I.

\subsection{ESO\,350-IG38 (\object{Haro 11})}

\begin{figure}
\resizebox{\hsize}{!}{\includegraphics{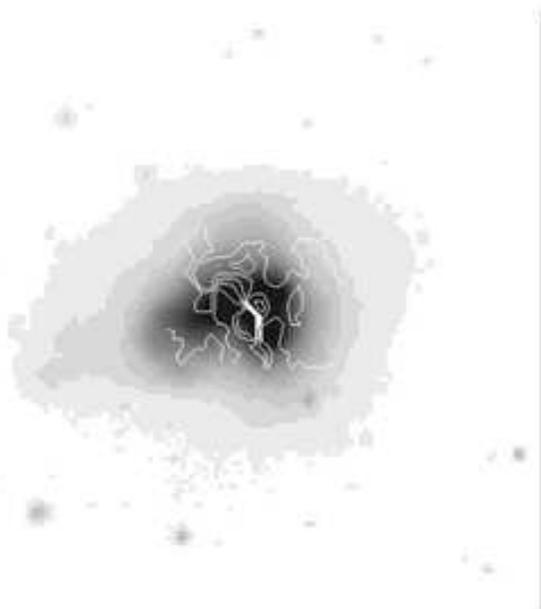}}
\caption[]{The H$\alpha$ iso-velocity contours overlayed on an R-band CCD 
image of ESO\,350-IG38. The velocities to which the contours correspond can
be found in Paper I. North is up, and east to the left. The size of the 
field is 1 by 1 arcminute, corresponding to $23 \times 23$ kpc. Note the irregular 
morphology at all isophotal levels. The faintest visible structures are 
$\mu_R \approx 26$ mag/arcsec$^2$. }
\label{contig38}
\end{figure}

This galaxy has a heart shaped morphology, with three bright starburst
nuclei (Fig. \ref{contig38}). The outer morphology is distorted
out to very faint isophotal levels 
(see Fig. \ref{contig38} where the faintest visible structures are 
$\mu_R \approx 26$ mag/arcsec$^2$). This is seen more clearly in broad
bands that in H$\alpha$ demonstrating that the light originates mainly in
stars. Hence, the large scale distribution
of stars in this galaxy is highly asymmetric. At all isophotal levels 
an extension in the south-eastern direction is evident.  This may be a 
tidal tail in development, or the remnants of a such.
H$\alpha$ images reveal faint arms extending from the south-eastern
and south-western nucleii.  
An HST/WFPC2 archive image  reveals that the three
very bright starburst nuclei at the center of ESO\,350-IG38 are
composed of numerous individual super-star clusters 
with luminosities up to  $M_V = -15$. The total
number of luminous star clusters is very high (\"Ostlin
\cite{ostlin:strasbourg}).  Emission from the south-east nucleus is
dominated by a resolved very high surface brightness ``nucleus'' with
$M_V = -17.5$ with no apparent internal structure.
The properties of these central regions  bear a close resemblance to
the classical colliding galaxies NGC 4038/4039 as seen in HST data  
(Whitmore and
Schweizer 1995).

The H$\alpha$ line profiles of this galaxy are broad, up to FWHM=270 km/s,
and have a non-gaussian shape. This suggests that two or more non-virialised
components may be present. In  the centre, double peaked lines are present 
consistent with the presence of a counter-rotating disk (see Fig. \ref{massmod1st4} and Paper I) or 
high velocity blobs. These properties indicate that the centre is not
dynamically relaxed, while the outer velocity field shows a very slow
rotation. The estimated stellar mass density exceeds by far what can
be supported by the observed amount of rotation ($\approx 30$ km/s). 
Thus, the galaxy is either not in equilibrium, or it is not primarily 
supported by rotation.  These properties strongly suggest that the starburst 
was triggered by a merger process. 

This galaxy is the most massive in
the whole sample and also has the highest star formation rate. The
inferred  (core-collapse) supernova frequency is one every 7 years.
Surprisingly, it seems to be rather devoid of cool gas (Bergvall et al. 2000), 
suggesting that the starburst is about to run out of fuel.

\begin{figure}
\resizebox{\hsize}{!}{\includegraphics{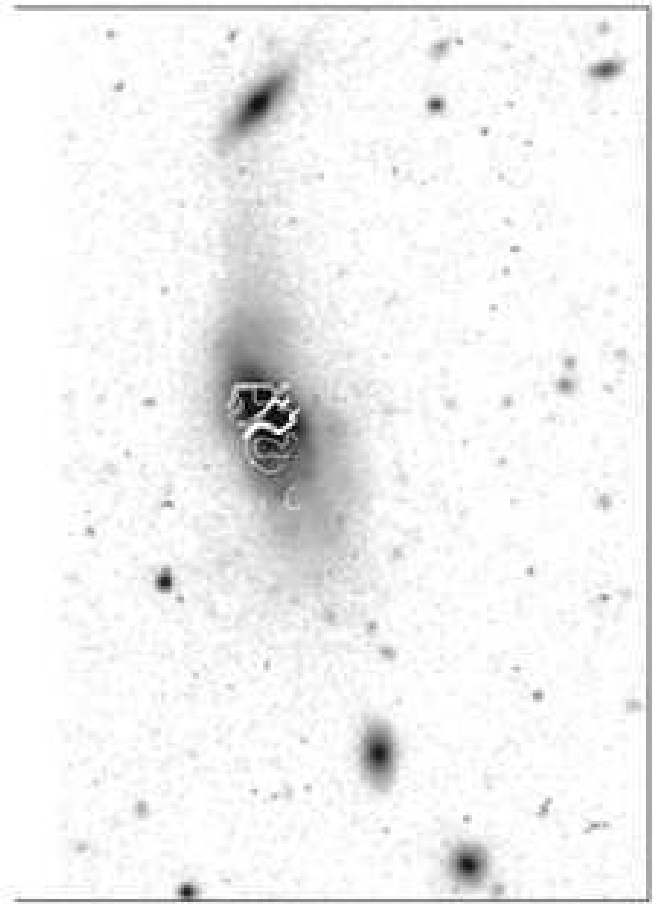}}
\caption[]{The H$\alpha$ iso-velocity contours of the primary dynamical
component in ESO\,480-IG12 (see Paper I) overlayed on an R-band CCD
image. North is up, and east to the left. The size of the field is 1.7
by 2.5 arcminutes, corresponding to $30 \times 45$ kpc.}
\label{contig12}
\end{figure}

\subsection{ESO\,480-IG12}

An intriguing finding with this galaxy is that it has strong outer
morphological distortions (Fig. \ref{contig12}) and that it is 
apparently aligned with a chain of galaxies in a background
(uncatalogued) cluster. Spectra have been obtained for three of the
galaxies with the smallest angular distance from ESO\,480-IG12, which
however are at much higher red-shift ($\sim 30000$ km/s), but it cannot be
excluded that one of the galaxies apparently belonging to the
cluster is in fact a low mass companion of ESO\,480-IG12.

The northern extension is long and narrow while the south one is
broader and more diffuse. South of the centre, there are several
kiloparsec scale plumes. In addition, there are several small faint
blobs in the southern part.  High resolution images would reveal if
these are extended or perhaps compact star clusters, like those seen in
ESO\,338-IG04, ESO\,350-IG38 and ESO\,185-IG13.
The overall morphology presents large-scale asymetries down to the 
faintest visible levels, compare e.g. the north and south-west 
extensions. This indicates an asymetric and non-equilibrium 
distribution of stars, which may be due to an interaction/merger
or a strong warp.

The velocity field is irregular with double line profiles, almost over
the entire galaxy, and two components have successfully been fitted in
Paper I. The double lines could arise in a separate
dynamical component or an outflow or expanding super bubble. The
rotation curve and luminosity distribution suggest that dark matter
dominates the dynamics (see Fig. \ref{massmod1st4}).

\begin{figure}
\resizebox{\hsize}{!}{\includegraphics{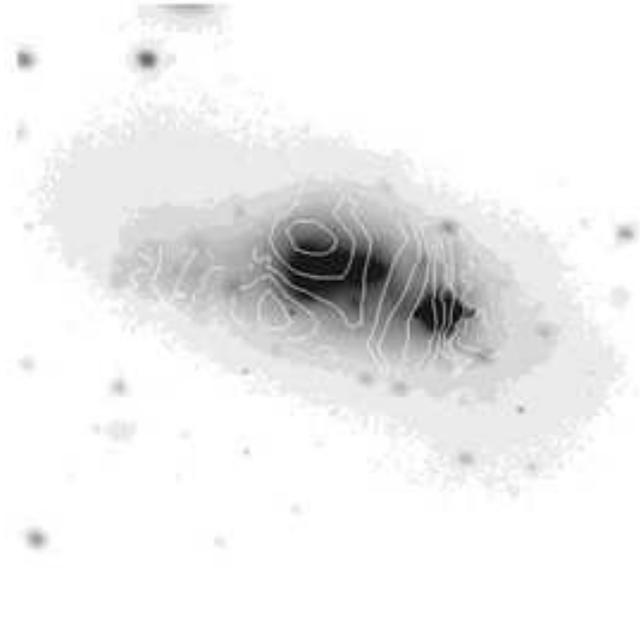}}
\caption[]{The H$\alpha$ iso-velocity contours of ESO\,338-IG04 overlayed on a 
R-band CCD image. North is up, and east to the left. The size of the field is 
$60\arcsec \times 53\arcsec$, corresponding to $11 \times 9.5$  kpc.}
\label{contig04a}
\end{figure}

\subsection{ESO\,338-IG04 (\object{Tololo 1924-416})}

This well-known BCG has an almost chaotic velocity field, with strong
gradients and an extended tail with little internal velocity structure
(Fig. \ref{contig04a}). It has a companion at a projected distance 
of 70 kpc to the south-west (see next subsection). Approximately 10\arcsec\ 
east of the centre, just at the border of the central star-forming region,
there is a velocity component whose kinematical axis is perpendicular
to the photometric major axis of the galaxy (se Paper I). 
 The radial light distribution indicates that
the observed mean rotational velocity (if a such is at all meaningful
to define in view of the irregular velocity field, see Fig. \ref{massmod1st4}) 
cannot support the
system gravitationally.

There is a 5 kpc long tail towards the east, and large scale isophotal 
assymetries down to the 26 mag/arcsec$^2$ level. At fainter levels the
morphology becomes more regular, but a box-like shape remains on 
the western side. The tail has much bluer colours 
than the rest of the galaxy outside the starburst region, signifying a
younger stellar population. It contains stellar clusters, and cannot be
explained by a purely gaseous tail. The tail has almost no velocity gradient 
with respect to the centre ($\Delta v \le 10$ km/s). On the other hand, the western half of the 
galaxy has a strong gradient and an implied rotational velocity of 80 km/s
at a distance of 3 kpc from the centre (Fig. 5 in Paper I). This is 
identical to the rotational velocity in  the companion ESO\,338-IG04B which 
has an equal photometric mass. Hence,  the western part of the galaxy 
shows about the expected level of velocity difference with respect to the
centre for rotational support to be possible. But what is happening at the 
eastern side in the tail? The  colour of the tail suggest that it has a distinctly
different stellar population from the rest of the galaxy. The most likely
is that the tail is a remnant of a merger and that projection effects
prevent us from seeing the true velocity amplitude. The parts of the galaxy
on the eastern side which are not in the tail, do not emit in H$\alpha$,
hence we have no information on its kinematics. Where the tail
meets the starburst region we see an increased H$\alpha$ velocity dispersion,
perhaps due to a shock. This coincides with the location of the perpendicular
dynamical component discussed above. The companion is probably too far away
for tidal forces to have caused the starburst and peculiar velocity field.

Radio interferometric observations (\"Ostlin et al. in preparation)
reveal that the galaxy is embedded in a very large H{\sc i} cloud, more than
7 arminutes across (corresponding to 80 kpc at the distance of
ESO\,338-IG04). The H{\sc i} cloud has irregular morphology with two main
components and no single axis of rotation.  ESO\,338-IG04 appears to be
located in the eastern H{\sc i} cloud.  The morphology and velocity field of
the H{\sc i} complex is consistent with a close interaction/merger of two
gas-rich galaxies or H{\sc i} clouds. The companion (see next subsection) 
is detected in H{\sc i} but lies further away.

Although we cannot exclude that the starburst in ESO\,338-IG04 is triggerred
by interaction with the companion, a merger appears more likely in view of
the complex velocity field and the non-rotating arm. HST observations of this 
well known starburst  has revealed that in addition to many  young compact
star clusters, it contains a system of intermediate age ($\sim 2$ Gyr) globular 
clusters (\"Ostlin et al. \cite{ostlin:e338gc}),  a fossil of a previous
dramatic starburst event.

\begin{figure}
\resizebox{\hsize}{!}{\includegraphics{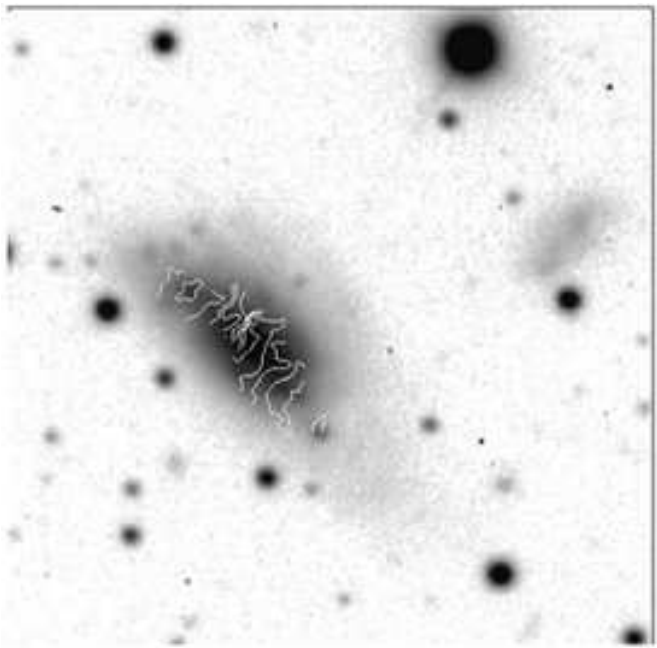}}
\caption[]{The H$\alpha$ iso-velocity contours of ESO\,338-IG04B overlayed 
on an V-band CCD image. North is up, and east to the left. The size of the 
field is $83\arcsec \times 83\arcsec$, corresponding to $15 \times 15$  kpc. Note the low 
surface brightness galaxy north-west of ESO\,338-IG04B}
\label{contig04b}
\end{figure}

\subsection{ESO\,338-IG04B (companion)}
This is a physical companion to ESO\,338-IG04.  
The projected distance between the galaxies is 70 kpc.
Although the colours are fairly blue, ESO\,338-IG04B is not a BCG, and the star
formation rate is moderate and does not imply a starburst (see Table 4).  
Photometry indicates an old underlying stellar
population (Bergvall and \"Ostlin 2000). The star formation in this
galaxy could have been enhanced by the tidal drag from ESO\,338-IG04,
but the velocity field is regular and apparently unperturbed
(Fig. \ref{contig04b}). The kinematics of this galaxy support the
presence of a dark matter halo (see Fig. \ref{massmod1st4}).  The stellar mass of this galaxy is of
the same order as that of ESO\,338-IG04, but the estimated SFRs 
differ by more than a factor of ten.  North-west of the galaxy a low surface brightness
galaxy is detected in broad band images, but not in the Fabry-Perot
H$\alpha$ images, hence its distance is unknown.

\begin{figure}
\resizebox{\hsize}{!}{\includegraphics{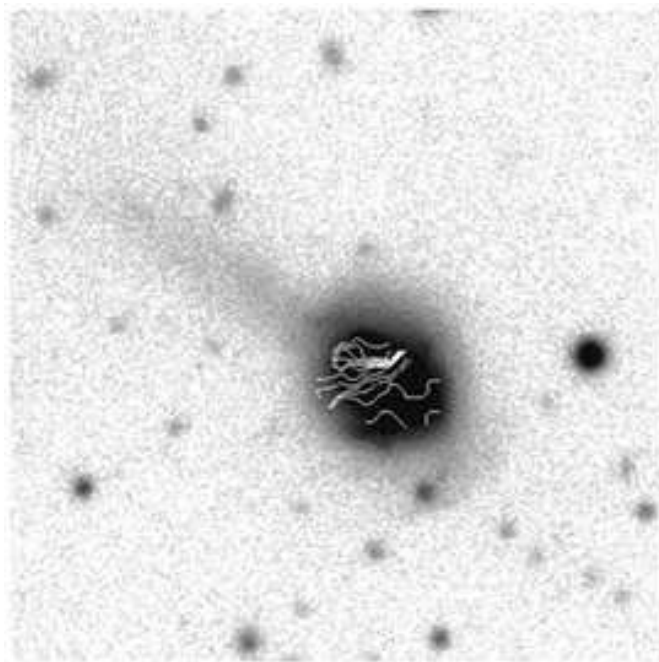}}
\caption[]{The H$\alpha$ iso-velocity contours of the primary dynamical 
component of ESO\,185-IG13 overlayed on a V-band CCD image. North is up, 
and east to the left. The size of the field is $50\arcsec \times 50\arcsec$  
corresponding to $18\times 18$ kpc.}
\label{contig13}
\end{figure}

\subsection{ESO\,185-IG13}

ESO\,185-IG13 presents a clear example of classical  morphological 
perturbations produced in galaxy mergers and/or collisions (Barnes and 
Hernquist \cite{barnes:hernquist}; 
Schweizer \cite{schweizer}). The broad band images of this galaxy (see fig. 10)
clearly reveal the presence of a tidal tail 30$\arcsec$ long in the north-
east direction (about 10 kpc at the distance of the galaxy main body).  

On CCD images, there are two small faint galaxies with fairly low surface
brightness approximately 1.5 arcminutes south-west of
ESO\,185-IG13. None of these were detected in the Fabry-Perot data,
hence their distances are unknown.  An HST image of
ESO\,185-IG13 (obtained from the archive) reveals the presence of
numerous compact star clusters with luminosities up to $M_V \approx
-15$ (\"Ostlin \cite{ostlin:strasbourg}),
a typical signature of colliding system (see Whitmore et al. 1993, 1995).  Most of
the bright cluster sources are concentrated to a central bar like
structure. Both the HST and ground based images reveal the presence
of arms in the center, though with no counterpart in the Fabry-Perot
$H\alpha$ image or velocity field.

The analysis of the velocity field of ESO\,185-IG13 revealed the presence
of two dynamically distinct components (Paper I).  The secondary component is
counter-rotating with respect to the main component (which spins
faster and dominates the H$\alpha$ emission). The two components have
marginally different position angles. If the two components represent
gas discs that intersect, this configuration cannot be long lived (the
rotation period of the secondary component is $\approx 4 \times
10^8$years). 
Despite a regular rotation curve, the observed level of rotation
cannot support the observed photometric mass (see Fig. \ref{massmod185}). However, given the 
limited surface photometry available, and the lack of near-IR data, the
estimated $M/L$ for the disk component is rather uncertain. If the
IMF has a flat low mass part (see Sect 2.4) and if we have underestimated
the amount of rotation (e.g. due to projection effects) and take into
account the sum of the two components (as in Fig. 2), it might be possible 
to solve the mass discrepancy. In summary, this is a clear example of a BCG where the 
starburst has been triggered by a merger.

\begin{figure}
\resizebox{\hsize}{!}{\includegraphics{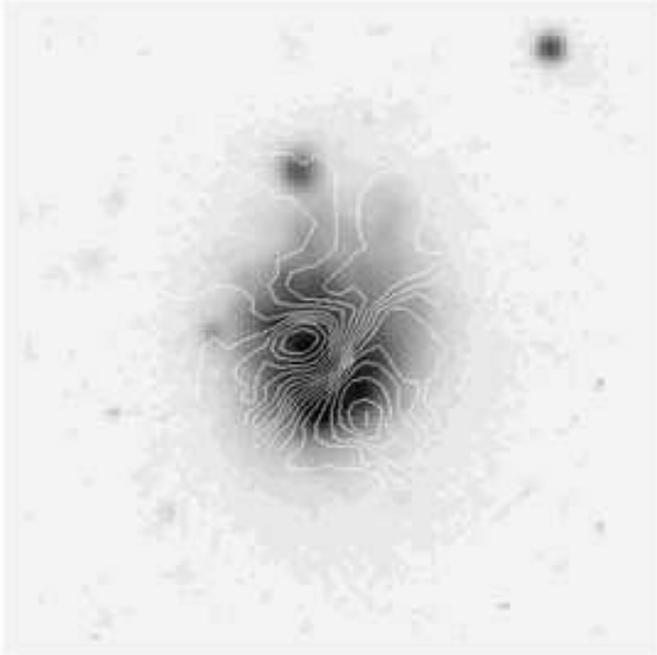}}
\caption[]{The H$\alpha$ iso-velocity of contours ESO\,400-G43 overlayed 
on an R-band CCD image. The grey scale have been chosen to show detail
rather than depth. North is up, and east to the left. The size of  
the field is $50\arcsec \times 50\arcsec$, corresponding to $19\times 
19$ kpc. The source 15\arcsec north of the centre is an H$\alpha$
emitting region at the red-shift of ESO\,400-G43.}
\label{contg43}
\end{figure}

\subsection{ESO\,400-G43}

This galaxy has a regular outer morphology, embedded in a large H{\sc i}
cloud that is extended towards a companion galaxy with a projected
distance of 70 kpc (Bergvall and J\"ors\"ater 1988, see next subsection).
At fainter isophotal levels peculiar morphological features start to appear.
Approximately 15\arcsec north of the centre there
is a bright H$\alpha$ emitting blob, apparently not rotating with the
galaxy at the pattern speed expected at its location.  

The rotation curve shows a rapid increase followed by a dramatic
decline, that drops faster than the keplerian prediction (see Fig. 
\ref{massmodlast4} and Paper I). This is obviously
unphysical for an equilibrium disc. A possible interpretation is that
the cause of the super-keplerian rotation speed is infall towards the
centre. A $B-R$ image of ESO\,400-G43 
shows a shell-like structure with
radius $\sim$1.5 to 3 kpc. This may be a sign of superbubbles or
stellar resonances, like the ``shells'' found in many merger remnants.
In the north-west, starting 10$\arcsec$ from center, there are signs
of an arm/plume both in the velocity field and broad band images
(Fig. \ref{contg43}).  Our kinematical data for the ionised gas is in 
good agreement with the results by Bergvall and J\"ors\"ater (1988). 
The published  H{\sc i} kinematical data show a rotation curve
with a rotational velocity of $\sim 50$ km/s (Bergvall and J\"ors\"ater 
1988), i.e. similar to the central value in the H$\alpha$ rotation curve 
before the super keplerian decline. Bergvall and J\"ors\"ater 
(1988) suggested that this galaxy was a genuinely young galaxy and
that the super-keplerian drop was due to a non-relaxed disk.
The latter conclusion remains likely, although it is evident that
the galaxy contains an old underlying population (Bergvall and \"Ostlin
2000).
 It is likely that a rapid infall of gas towards the center is  fuel
 for the starburst but -- what gave rise to the infall? The most efficient infall
mechanisms are bar instabilities and mergers, but there are no signs
of a bar in ESO\,400-G43.  The northern bright H{\sc ii} region might be an
infalling blob, or something that has just plunged through the H{\sc i}
disc. Slightly more than an arcminute to the north-east there is a
positive detection of a massive H{\sc i} cloud, which may be interacting
with ESO\,400-G43 (Bergvall and J\"ors\"ater 1988).

The low rotational velocity 
outside the centre cannot, by far,
support the stellar mass gravitationally (Figs 2 and 5).  The observed H$\alpha$ line
widths (see Table 4) could be  sufficient for
gravitational support if the linewidth traces virial
motions. However, this can not explain the observed shape of the
rotation curve. Clearly, this galaxy is badly perturbed dynamically. The only
likely explanation is interaction with the companion ESO\,400-G43B
or that a merger with a smaller galaxy has occurred.

\begin{figure}
\resizebox{\hsize}{!}{\includegraphics{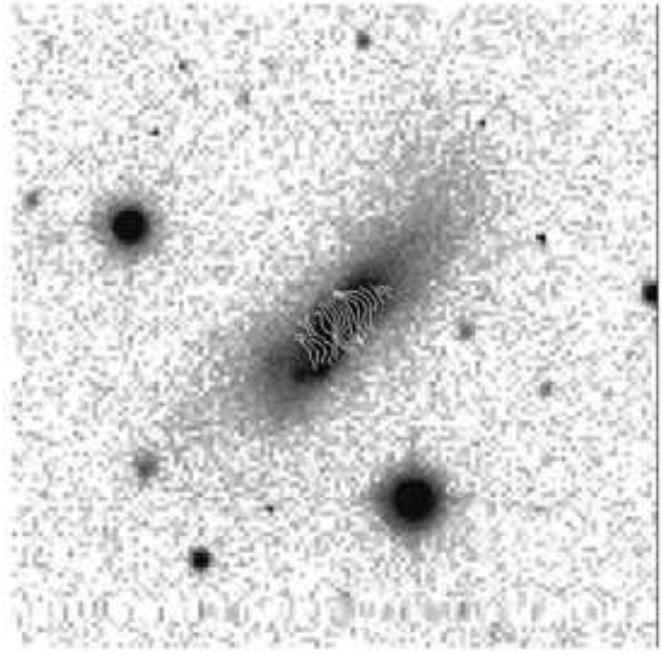}}
\caption[]{The H$\alpha$ iso-velocity contours of ESO\,400-G43B
overlayed on an R-band CCD image. North is up, and east to the
left. The size of the field is 1 by 1 arcminute, corresponding to 
$22 \times 22$ kpc. }
\label{contg43b}
\end{figure}

\subsection{ESO\,400-G43B (companion)}

This physical companion  to ESO\,400-G43 (Bergvall and
J\"ors\"ater 1988)  is a regular dwarf galaxy with a regular rotation curve
(Figs. \ref{massmodlast4} and \ref{contg43b}). The shape of the rotation
curve does not follow the light distribution which indicates the presence of a
dark matter halo (Fig. \ref{massmodlast4}). The properties of ESO\,400-G43B are
fairly similar to those of ESO\,338-IG04B, but it
has more intense star formation activity. The time scales for gas
consumption and buildup of the observed stellar mass are both on the order of
3-4 Gyr.  Thus, this galaxy is at the limit of being classified as a  starburst
galaxy.  The star formation occurs in a central extended HII
region. It is quite probable that interaction with ESO\,400-G43 
increased the star formation activity.

\begin{figure}
\resizebox{\hsize}{!}{\includegraphics{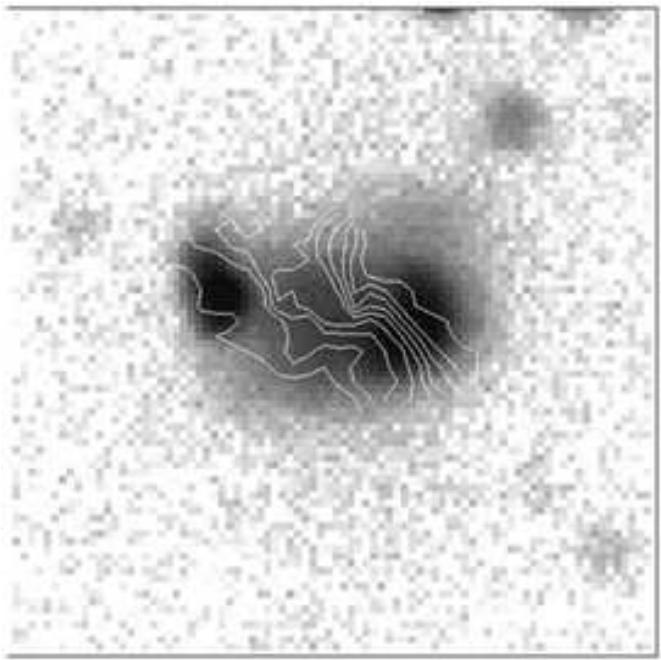}}
\caption[]{The H$\alpha$ iso-velocity contours of Tololo\,0341-407
overlayed on an R-band CCD image. North is up, and east to the
left. The size of the field is $37\arcsec \times 37\arcsec$, 
corresponding to $11 \times 11$  kpc.}
\label{conttol}
\end{figure}

\subsection{Tololo\,0341-407}

This galaxy is the faintest BCG in the sample. It has two apparent nucleii 
embedded in an irregular envelope
(Fig. \ref{conttol}).  The velocity field shows the two nuclei to
belong to distinct dynamical systems.  In Paper I we successfully
decomposed this galaxy into two dynamical components coinciding with
the optical components on the broad band image.  This may be two dwarf 
galaxies currently coming together. The eastern component (Tololo\,0341-407E) 
has a very high SFR per unit mass, whereas Tololo\,0341-407W is similar
to the more luminous BCGs in this respect. Notice that both galaxies
are far less massive than the other galaxies in the sample. 

This is the only galaxy in the sample for which we do not have
multi-colour CCD photometry, hence we could not determine $M/L$. As an
estimate for $M/L$ we took the median for the other galaxies, and for
$M/L_{\rm disk}^{\rm min}$ and $M/L_{\rm disk}^{\rm max}$ we used the extreme values of the
distribution for the other galaxies. Hence, the mass estimates for this
system are more uncertain than for the other galaxies. Nevertheless, 
the maximum disk solutions are close to the predicted photometric 
rotation curves, and in the Eastern component we see marginal evidence
for a dark matter halo (Fig. \ref{massmodlast4}).


\section{Discussion}

\subsection{Energetic feedback from star formation}

Supernova (SN) explosions and winds from massive stars provide input of 
mechanical energy into  the ISM in galaxies. In a starburst, the 
collective action of many supernovae may lead to a galactic wind, 
transporting material out from the plane of the galaxy. Winds could
affect the velocity field of the ionised gas and has to be taken 
into account. Observationally, winds have been seen in some low luminosity, 
hence low mass, BCGs (Meurer et al. \cite{meurer}, Papaderos et al. 
\cite{papaderos},  Lequeux et al. \cite{lequeux95}, Martin \cite{martin}).
 
Recent models which include also the restoring force provided by dark 
matter have shown that feedback from star formation is not expected to 
cause gas loss from galaxies more massive than $10^8 M_\odot$ (MacLow and 
Ferarra 1999, Ferarra and Tolstoy 2000). Thus for the galaxies studied in 
this paper, it should be safe to 
conclude that feedback has not affected the global velocity fields. However, 
the calculations by MacLow and Ferarra (1999) and  Ferarra and Tolstoy (2000) 
are restricted to energy injection rates 
corresponding to supernova frequencies of one SN every 30 000 years or less.
In this sample, where the derived star formation rates span the range  0.25 
to 20  $M_\odot$/year,  we expect (assuming a Salpeter IMF with 
$M_{\rm low} = 0.1$, $M_{\rm up} = 100 M_{\odot}$,  and lower initial mass 
limit for core-collapse supernova of 8  $M_{\odot}$) SN frequencies of 0.02 to 
0.15 SN/year, hence 3 orders of magnitude larger! Extrapolating the calculations 
by MacLow and Ferarra we still do not expect blow-away to occur, while blow-out 
is possible. Of course, extrapolating by 3 orders of magnitude is dubious. 
If gas gets  blown out, it will stay in the halo and may later condense back 
on the host galaxy. 

Interestingly, deep H$\alpha$ images of ESO\,338-IG04 and ESO\,350-IG38 show that 
the H$\alpha$ morphologies are extended along the minor axis, compared to 
the broad band morphologies. 
This is what is expected for disk galaxies: gas escapes perpendicular to 
the plane while the gas in the plane is little affected. Thus the 
rotation curves, which are derived along the major axis, are probably not 
affected by ouflows and expanding superbubbles  while the velocity fields 
along the minor axis may well be (e.g. in ESO480-IG12).

\subsection{Morphology}

Our sample of relatively luminous BCGs have irregular morphology at both
bright and faint isophotal levels. This agrees with the findings of 
Telles et al. (\cite{telles:etal}) that BCGs with irregular morphology at 
faint isophotal levels  are on average more luminous than those with regular 
morphology.

A galaxy in equilibrium should have a regular symmetric morphology and
kinematics.  The occurence of star formation, if not symmetrically
distributed, may create apparent morphological irregularity. Nevertheless, 
the outer parts of a galaxy are
essentially unaffected by the central activity, hence the outer
isophotes of a galaxy mainly trace the distribution of
stars. Therefore, significant large scale asymetries in the outer isophotes are
signatures of an asymetric distribution of stars.  Such asymetries
could be the consequence of a interaction/merger but weaker distortions
could also be produced by instabilities which may be inherent or tidally 
triggered. 
Mergers and interactions between galaxies  lead to
large scale morphological distortions (e.g. tidal tails and shells) 
and to perturbations in their rotation curves due to  gas flows driven 
into the center.
The light distributions of the 6 BCGs presented in this study show outer
distortions to some extent which may suggest  that the starbursts  have 
been produced by interactions or mergers.  In some cases, like ESO\,185-IG13 
and ESO\,350-IG38, the merger signatures are very strong.

In principle, one of the best ways to investigate the effect of
interaction between galaxies should be through the kinematics of the
gas, since it traces  well the dynamics of the burst tidally
triggered trough the collision of gas-rich systems. From the reported
results we have shown that on the top of a slow rotating disk system,
important dynamical perturbations have been detected which prevent us
in some cases from obtaining the gravitational mass of the galaxies. None
of the BCGs show  regular disk kinematics and different explanations
have to be invoked for each particular case (see Sect. 3 and 4).  It has to
be noticed that the only truly regular kinematics occur  in the
companion galaxies or when we can disentangle quite well the different
kinematical components, as in the cases of ESO\,185-IG13, ESO\,480-IG12 and
Tololo\,0341-403. It is then not surprising that we cannot define a general trend 
for all the studied galaxies. Depending on the state in which we see the 
interaction/merger, we will detect more or less chaotic velocity maps.

\subsection{Environmental aspects} 

Most investigations of the environments of BCGs have shown that these
kind of galaxies are rather isolated with respect to giant luminous
galaxies and preferentially occupy low and intermediate density 
environments (see Introduction).
We have used NED\footnote{NASA/IPAC Extragalactic Database} to 
investigate the environment of the galaxies in
this study.  Since NED was mainly composed from catalogues of bright
galaxies, such a study essentially provides information on the
presence of massive neighbouring galaxies.  We searched a projected 
radius of 1 Mpc around the galaxies, and $\pm 1000$ km/s in velocity space.  
The velocity range was chosen very generously so as not to reject possible neighbours
with high peculiar velocities. Travelling one Mpc with a
velocity of 1000 km/s would take one Gyr. Thus, if galaxies so far
apart  have ever interacted, this must have happened more than 1 Gyr
ago, even for very high peculiar velocities. Hence, any signs of interaction
(e.g. a triggered starburst) should be long gone. 
None of the galaxies in this study has any neighbour in
NED within 1 Mpc, except for  ESO\,338-IG04 and ESO\,400-G43, the companions
of which are included in this study.  

The analysis shows that the local density of catalogued galaxies is not
enhanced in the locations where these galaxies are found; typically
we find a density of catalogued galaxies of 0.005 Mpc$^{-3}$, when
averaged over a volume 2500 Mpc$^{3}$ or greater. 
Thus, the sample studied is not located in particularly dense 
environments. 
In general, galaxy evolution is a strong function of the density of the 
environment. In low red-shift clusters, interactions are frequent but
rarely lead to any significant star formation. The most active star-forming 
galaxies and star-forming mergers are found in the field or
in sparse groups, and this is true also for BCGs, luminous or not.

\subsection{The ignition of a starburst}

A fundamental astrophysical problem is how to ignite a starburst. The 
majority of galaxies are not involved in starbursts. For example, low
surface brightness galaxies (LSBGs) have very low star formation rates, 
despite the generous supply of H{\sc i}, which may be explained by 
sub-critical H{\sc i} column densities (van der Hulst 
et al.  1993). Efficient star formation requires gas with high densities. 
Here we consider as starbursts, galaxies which have gas depletion time-scales, 
and time-scales for accumulating the observed photometric mass with 
the current SFR much shorter than a Hubble-time, i.e. $\tau_{\rm ph} \sim 1$
Gyr or less.

A model where starbursts in dwarf galaxies are the consequense of
statistical fluctuations was presented by Gerola et al. (1980). Their
model predict that for galaxies with radii less than 1-2 kpc, star
formation will principally occur in short bursts, whereas larger
galaxies are predicted to have more continous star formation. 
A similar argument is that only when
the star formation in a single giant molecular cloud would exceed the
time averaged SFR in a galaxy would we expect statistical fluctuations
to dominate. Hence for galaxies with a baryonic mass greater than $10^8 M_\odot$,
statistical fluctuations should not dominate the global SFR. The galaxies 
presented in this paper  are too large/massive for any significant  
statistical fluctuations to be expected.


A popular scenario for BCGs has been one with cyclic bursts. Here,  the 
starburst is terminated by the expulsion of gas through supernova winds. 
If the gas later accretes back to the galaxy, a new starburst could be 
ignited.  A potential problem is, however, the time-scale for the retrieval 
of the gas. If the blown out gas eventually accretes back on the galaxy, one 
expects this to occur over a time-scale much longer than the duration of 
a starburst.  
A constant  inflow of gas at a slow rate is  not likely to 
produce a starburst, unless there is some threshold mechanism. 
But even if there is a treshold, the accreted gas mass has to pass it coevally. 
Since the star formation efficiency is less than 100\%, the gas accretion 
rate must be higher than the SFR. Such a high continous mass accretion rate 
would not be compatible with the total gas+star masses of the galaxies.
For our BCGs which have  burst masses of $10^8 M_\odot$ or larger, 
on the order of a few times $10^8$ to $10^9 M_\odot$ of gas would have to pass the treshold 
within a few times $10^7$ years. This implies a mass accretion time 
scale of the same order as the free fall time.
Hence, gas falling back would have to come all at once in one lump,
contrary to the expected continous accretion. One could of course envision 
that  the expelled gas cools and clump together in a few giant massive 
H{\sc i} complexes, and if such an aggregation would fall into the centre, 
we could get a starburst. 
The time-scale for building up the observed photometric mass  with the 
current SFR is $\tau_{\rm ph} \sim 1$ Gyr (see Table 4), which indicates 
that if the burst duration is $\le 100$ Myr, the duty cycle between bursts 
should be on the order of 1 Gyr. We would thus require that gas is 
ejected on time scales shorter than 100 Myr,  then the gas should stay away 
for 1 Gyr, and finally, most of it should fall back within some 10 Myrs. 
Supernova activity will begin after a few Myrs, whereas the bursts we observe
in these BCGs have already been active for several 10 Myrs. Hence, although we 
in some cases see signatures of expanding bubbles,  we see no evidence
for any superwinds capable of blowing out the ISM and terminating star formation.
Because of these arguments, we do not consider a self-regulated cyclic starburst 
scenario very likely for these luminous BCGs. A cyclical burst scenario, 
powered by winds or stochastic effects, is more likely in low mass BCGs.


Tidal forces generate gas flows that could funnel gas at large radii into
the  center of a galaxy and ignite a starburst, but generally on time scales 
that are too  long to build up high central densities in competition 
with the destruction processes of molecular cloud complexes. 
Campos-Aguilar et al. (\cite{campos:1993}) showed that a dwarf 
galaxy is sensitive to tidal triggering only when the companion 
galaxy is very massive. With the lack of massive companions, interactions 
will only be strong  when two dwarfs are actually merging or in physical 
contact. Similarly, Icke (1985) found that separations between galaxies
for tidal gas flows to occur must be within a few galactic radii, but  
the typical time-scale is of the order of a galactic rotation period 
(here a few 100 Myrs). Thus, a weak tidal gas flow might well increase 
the star formation rate but likely not ignite a starburst. Mihos et al. (1997) used
N-body simulations to study the response of LSBGs (possible progenitors of
BCGs) to tidal interactions. They found that due to the low mass densities, 
LSBGs are very stable against tidal triggering, in contrast to normal 
high surface brightness  giant disc galaxies.   As there are no massive
galaxies in the vicinity of our studied BCGs, and the two known 
companions are rather distant (more than 70 kpc), tidal interactions 
cannot be the main cause of the starbursts. However, tidal interactions
may be responsible for the moderately high SFRs seen in the two companions.

Another scenario for BCGs is that the starburst is ignited through the
merging between two dwarfs, or between a dwarf and a massive
intergalactic gas cloud.  In a merger, gas clouds lose a considerably 
fraction of their angular momentum in  dissipative collisions and fall 
into the center of the galaxy/galaxies. The time-scale is sufficiently short
for building up large overdensities that can start to form stars
coevally. This is thus an attractive scenario for the origin of
starbursts in BCGs. Massive isolated H{\sc i} clouds seem to be very rare
at the current epoch (see Kunth and \"Ostlin 2000). However, any merger 
involving a sufficiently gas-rich component could do the job.  Mergers 
involving at least one gas-rich component is the mechanism favoured for 
the  BCGs in this sample.

HST observations of ESO\,338-IG04 has revealed that it contains a rich 
population of young globular clusters (\"Ostlin et al 1998). Similar
rich populations of luminous star clusters were found in HST archive 
images of ESO\,350-IG38 and ESO\,185-IG13 (\"Ostlin  \cite{ostlin:strasbourg}).  
Numerous young globular cluster candidates have also been found in mergers, 
like the "Antennae" and NGC 7252 (Whitmore et al. 1993, 1995), and
galactic bars (e.g. Barth et al. 1995). In the latter case, the star 
cluster formation probably is triggered by resonances enhancing the gas density.
With respect to the host galaxy luminosity, luminous BCGs like  
ESO\,338-IG04, ESO\,350-IG38 and ESO\,185-IG13 contain higher
numbers of compact star clusters than the classical mergers  (\"Ostlin  
\cite{ostlin:strasbourg}).
A nearly coeval formation of a hundred globular clusters 
requires the build up of massive dense gas clouds on a time-scale of a 
few 10 Myrs. In dwarfs, where resonances that could help build-up 
large overdensities, are absent, mergers may be the only way to create 
the necessary conditions for cluster formation to occur in great numbers
(see e.g. Elmegreen and Efremov 1997). Hence the richness of young 
massive star clusters in three of our galaxies is another support for 
the merger origin of their starbursts.

\subsection{Relations to other galaxy types}

Given that the high SFRs seen in these luminous BCGs are transient,
one might ask what came before, and what will follow: in between the 
burst phases a BCG will have a very different appeareance. It is  
however hazardeous to draw any general conclusions about the BCG 
pehomenon from such a small sample of luminous BCGs. The morphological 
diversity, notably the existence of BCGs with regular vs irregular outer 
envelopes, suggests the possibility that different types of BCGs may 
have a different origin (Kunth and \"Ostlin 2000). 

The possible evolutionary links between BCGs and other types of galaxies 
have been widely discussed in the litterature (see e.g. Davies and Phillips
1988; Dekel and Silk 1986; Lin and Faber 1983;  Thuan 1992, Babul and Rees 
1992; Papaderos et al. 1996; Bergvall et al. 1998,1999; Kunth and \"Ostlin 
2000). For the luminous BCGs in this study, a merger origin is favoured for 
most of the galaxies. The merger should include at least one gas-rich 
dwarf or alternatively massive intergalactic H{\sc i}-cloud. This 
suggests that at least one dwarf irregular or LSBG is involved. 
LSBGs  may  be the most common type of galaxy 
in the universe, thus providing plenty of fuel for BCG activity.
Mergers between spirals are believed to produce elliptical galaxies. Since 
several of our galaxies seem to rotate too slowly, one might speculate that 
they will evolve into low mass elliptical galaxies. However, to answer the 
question about the fate of these BCGs,  information on  stellar dynamics 
is required.

Studies of faint galaxies, e.g. in the Hubble Deep Field (HDF), has revealed 
the presence of a population of high red-shift ($z = 0.4$ to $1.4$) compact 
emission line galaxies (Phillips et al. 1997),
which may be responsible for up to 45\% of the global cosmic SFR at $0.4 < z < 1.0$
(Guzm\'an et al. \cite{guzman}).
Comparing the mass averaged SFRs, line widths, line ratios etc, of our local
BCGs to those derived for the compact galaxies in the HDF (Guzm\'an et al. \cite{guzman}),
we find them to have very similar properties, in particular when compared 
to the high $(z>0.7)$ red-shift subsample.
Glazebrook et al. (1995,1998) identified the excess of faint blue galaxies
in deep galaxy counts with peculiar/interacting galaxies, and showed that
such galaxies are blue, compact and actively star-forming, with typical 
luminosities  $M_B = -19$, i.e. very similar to the galaxies discussed in 
this paper. 
In  hierarchial galaxy formation scenarios, massive galaxies are 
succesively built up by galaxies of lower mass. Le F\`evre et al. (\cite{lefevre}) 
show that mergers play a major role in  galaxy evolution also at $z<1$.
Thus, the local luminous BCGs  may be the present day equivalents (NB not counterparts) 
of high red-shift BCGs and faint blue galaxies.
The study of local luminous BCGs offer insights to the mechanisms operating
at higher red-shifts and offers a way of studying hierarchial buildup in action.

\section{Summary and conclusions}

We present the results from the most extensive study as yet of optical
2D velocity fields of luminous blue compact galaxies (BCGs) utilising 
Fabry-Perot interferometry, targeting the H$\alpha$ emission line. 
The velocity fields and rotation curves for a sample of six luminous BCGs 
and two companions were presented in a previous paper (\"Ostlin et al. 1999, 
Paper I). The velocity fields present large scale peculiarities, and secondary 
dynamical components (e.g. counter-rotation) are common.

In this paper we have analysed the ionised gas dynamics together with 
optical/near-infrared surface photometry and spectral synthesis models to 
constrain the dynamical and photometric masses of the galaxies. Moreover, 
we construct multi-component mass models including a dark halo component.   

The BCGs in this study are starbursts in the sense that the time-scale for 
building up the observed stellar mass with the current star formation rate,
derived from the H$\alpha$ luminosities, is much smaller than a Hubble time. 
We find that the young burst population dominates the integrated optical
luminosities, while only contributing 1 to 5\% of the total stellar mass, 
which ranges from a few times $10^8$ to more than  $10^{10} M_\odot$. The
mass is dominated by an older underlying population and the integrated 
(burst + old population)  mass-to-light ratios are $M/L_V \sim 1$. 

In about half the cases, the observed rotational velocities are too small 
to allow for pure rotational support. A possible explanation is that velocity 
dispersion dominates the gravitational support. This is consistent with the 
observed line widths ($\sigma_{{\rm H}\alpha} = $ 35 to 80 km/s), but does
not explain the strange shape of many of the rotation curves. 
Another possibility is that the galaxies are not in dynamical equilibrium, 
e.g. because they are involved in mergers, explaining the peculiar kinematics. 
It is also possible that gas and stars are dynamically
decoupled and the H$\alpha$ velocity field does not trace the gravitational potential. 
A way to distinguish between these alternatives would be to obtain the rotation 
curve and velocity dispersion  for the stellar component.
In two cases, we find evidence for the presence of dark matter within
the extent of the  H$\alpha$ rotation curves, and in two other cases we find marginal
evidence.

We have further analysed the morphology of the BCGs, and in general we
find strong large-scale asymmetries down to the faintest isophotal levels, 
revealing large scale asymmetries in the distribution of stars. In most 
cases we see clear signatures of merging/interaction.

We have discussed the possible trigger mechanisms for the strong starbursts 
in this sample of luminous BCGs. When considering also the kinematics and 
morphologies we are lead to the conclusion that dwarf galaxy mergers is the 
favoured explanation for the starbursts. The two companion galaxies appear largely 
unaffected by the presence of their brilliant neighbours, but their star 
formation activity may have increased due to the tidal drag imposed by the BCGs. 
The dynamics of 
the studied galaxies fall in two broad classes: one with well-behaving rotation 
curves at large radii and one with very perturbed dynamics. This may indicate a 
distinction of the fate of these galaxies, once the starbursts fade. Alternatively,
depending on the state in which we see the interaction/merger, we will detect more 
or less chaotic velocity fields. These local 
dwarf galaxy mergers may be the best analogues of hierarchical 
buildup of more massive galaxies at high red-shifts.

In 1999 and 2000 new Fabry-Perot observations were obtained for another
15 BCGs, extending down to fainter luminosities. Together with the
present data set, we will have high quality velocity fields for two dozen
BCGs. This will allow us to obtain a more comprehensive picture of
the evolution of BCGs.

\begin{acknowledgements}

Thanks to Torgny Karlsson for  reducing a CCD image of ESO\,185-IG13,
and to Ernst van Groningen for obtaining additional data on this
galaxy. Thanks also to Mats Dahlgren for obtaining the CCD image of Tololo\,0341-407. 
Claude Carignan is thanked for fruitful discussions on the dynamics of 
BCGs and for making
specific adaptations of his mass model code for the present study.
Albert Bosma is thanked for stimulating discussions on the interpretation 
of velocity fields. Anna Westman is
thanked for her great help in preparing the figures. An anonymous referee
is thanked for many useful comments on the manuscript.  G. \"Ostlin
acknowledges financial support from the Swedish Foundation for
International Cooperation in Research and Higher Education (STINT).
This work was partly supported by the Swedish Natural Science Research
Council. This work is based on observations collected at the
European Southern Observatory La Silla, Chile.

\end{acknowledgements}

\end{document}